\documentclass[twocolumn,showpacs]{revtex4}
\usepackage{graphics}

\begin{document}

\title{Spatial properties of entangled photon pairs generated in nonlinear layered structures}

\author{Jan Pe\v{r}ina, Jr.}
\affiliation{Palack\'{y} University, RCPTM, Joint Laboratory of
Optics, 17. listopadu 12, 771 46 Olomouc, Czech Republic}
\email{perinaj@prfnw.upol.cz}

\begin{abstract}
A spatial quantum model of spontaneous parametric down-conversion
in nonlinear layered structures is developed expanding the
interacting vectorial fields into monochromatic plane waves. A
two-photon spectral amplitude depending on the signal- and
idler-field frequencies and propagation directions is used to
derive transverse profiles of the emitted fields as well as their
spatial correlations. Intensity spatial profiles and their spatial
correlations are mainly determined by the positions of
transmission peaks formed in these structures with photonic bands.
A method for geometry optimization of the structures with respect
to efficiency of the nonlinear process is suggested. Several
structures composed of GaN/AlN layers are analyzed as typical
examples. They allow the generation of photon pairs correlated in
several emission directions. Photon-pair generation rates
increasing better than the second power of the number of layers
can be reached. Also structures efficiently generated photon pairs
showing anti-bunching and anti-coalescence can be obtained. Three
reasons for splitting the correlated area in photonic-band-gap
structures are revealed: zig-zag movement of photons inside the
structure, spatial symmetry and polarization-dependent properties.
Also spectral splitting can be observed in these structures.
\end{abstract}




\maketitle

\section{Introduction}

The first temporal correlations between the signal and idler
photons generated in the process of spontaneous parametric
down-conversion (SPDC) were observed already more than thirty
years ago \cite{Hong1987}. They manifested quantum entanglement in
the common state of both photons generated in one quantum event of
the spontaneous nonlinear process \cite{Mandel1995}. Since then
the understanding of properties of such photon pairs has grown
enormously. Polarization properties and namely polarization
entanglement between the signal and idler photons originating in
tensorial character of the nonlinear susceptibility have attracted
the greatest attention from the very beginning. The reason lies in
the simplicity of such states described in the Hilbert space with
dimension $ 2 \times 2 $. This made these states easily
experimentally accessible. Despite their simplicity, such states
have allowed to reveal many substantial features of quantum
physics related to correlations between subsystems
\cite{Bouwmeester2000}. Violation of the Bell inequalities that
ruled out neoclassical theories \cite{Perina1994}, the effect of
collapse of a wave-function \cite{Perina1994} and the ability of
teleportation of a quantum state \cite{Bouwmeester1997} belong to
the most important.

In the time domain, both photons occur in a very narrow temporal
window \cite{Hong1987} because they are emitted at one instant
after the annihilation of a pump photon. Moreover the conservation
law of energy dictates entanglement among monochromatic components
of the signal and idler fields \cite{Mandel1995}. This
entanglement leads to a typical finite entanglement time between
the detection instants of the signal and idler photons observable
in a Hong-Ou-Mandel interferometer \cite{Hong1987} or
sum-frequency generation of the paired photons
\cite{Harris2007,Brida2009}. These properties have been
extensively studied namely in connection with the generation by
pulsed pump fields \cite{Keller1997,PerinaJr1999a} that allows
precise synchronization of photons from different photon pairs.

Spatial properties of photon pairs have attracted attention last.
Correlations in the transverse planes of the signal and idler
photons occur here due to geometric properties of a photon-pair
source and pump-beam spatial profile. They originate in the
necessity of reaching good spatial phase matching of the
interacting fields \cite{Mandel1995} that results in an efficient
nonlinear process. For example, the sum of wave vectors of the
signal and idler fields has to approximately give the wave vector
of the pump field to observe an efficient photon-pair generation
in an extended bulk crystal using collimated pumping. This results
in strong correlations in emission directions of the signal and
idler photons \cite{Jobeur1994,Jobeur1996,Vallone2007}. These
correlations may even be exploited to 'transfer' spatial
properties of the pump beam into spatial correlations of the
signal and idler beams \cite{Monken1998,Walborn2004}. We may
observe analogy between spatial and spectral correlations
(entanglement) of two photons in a pair. Similarly as spectral
correlations may be tailored by a pump-field spectral profile,
spatial correlations may be controlled by a pump-beam spatial
profile. Also signal and idler electric-field phase variations in
the transverse plane belong to important characteristics of photon
pairs. They can be quantified in terms of eigenstates of the
angular orbital momentum operator \cite{Law2004}. Under certain
conditions, entanglement between such states in the signal and
idler fields has been observed \cite{Mair2001,Oemrawsingh2005}.
Experimentally, spatial correlations can be conveniently measured
by moving a fiber tip in the transverse plane, as it has been
done, e.g., in \cite{Molina-Terriza2005}. The use of an
intensified CCD camera has provided a more elegant way for the
experimental investigations of spatial correlations
\cite{Jost1998,Haderka2005}. Spatial correlations between the
down-converted beams have also been extensively used for quantum
ghost imaging \cite{Rubin2008}. We note that spatial correlations
occur not only in spontaneous regime, they have been found also in
the regime with prevailing stimulated emission
\cite{Brambilla2004,Jedrkiewicz2004}.

All these forms of entanglement are potentially interesting both
for fundamental physical experiments and practical applications
including metrology \cite{Migdall1999}, quantum cryptography
\cite{Lutkenhaus2000} and quantum-information processing
\cite{Bouwmeester2000}. In principle, all these forms of
entanglement can occur simultaneously depending on the source of
photon pairs. However, usual sources of photon pairs are
constructed such that only one or two forms of entanglement (e.g.,
polarization and spectral) are found and may efficiently be
experimentally exploited.

The effort to generate entangled photon pairs extended over as
many basis states as possible (and defined in several degrees of
freedom) and to modify 'the structure of entanglement' belong
together with the effort to enhance quantum efficiency of the
generation process to the leitmotifs of the development in this
field. Bulk nonlinear crystals that were nearly exclusively used
at the beginning have been gradually replaced by more complex and
efficient nonlinear structures including poled nonlinear materials
\cite{Harris2007,Kitaeva2007,Svozilik2009,Svozilik2010}, nonlinear
waveguides \cite{URen2004,Spillane2007,Chen2009} and nonlinear
photonic structures \cite{Bertolotti2001}. Nonlinear photonic
structures are extraordinarily interesting because they allow
efficient photon-pair generation owing to enhanced electric-field
amplitudes observed inside structures with photonic band-gaps
\cite{Yablonovitch1987,John1987} on one side, they also allow
relatively wide tailoring of properties of the emitted photon
pairs on the other side
\cite{Vamivakas2004,Centini2005,PerinaJr2006}. Structured
nonlinear fibers that rely on four-wave mixing
\cite{Li2005,Fulconis2005,Fan2005} represent a typical example. Or
waveguides with Bragg mirrors \cite{Abolghasem2009,Svozilik2011}
can be mentioned as perspective sources.

In the article, we consider systems composed of parallel nonlinear
layers. Back-scattering of the fields on the boundaries between
layers provides suitable conditions for the enhancement of
electric-field amplitudes under certain conditions \cite{Yeh1988}.
Moreover, spatial properties of photon pairs can be efficiently
tailored using parameters of these systems (e.g., the number of
layers). Nonlinear layered structures have already been studied in
the simplified geometry both in the framework of classical
\cite{Centini2005} and quantum \cite{PerinaJr2006} descriptions.
Here, we generalize the quantum model present in
\cite{PerinaJr2006} to the real spatial geometry including
vectorial character of the interacting fields. This allows us to
study transverse intensity profiles of the down-converted beams as
well as correlated areas of the signal and idler photons
considering typical layered structures made of GaN/AlN. We note
that GaN/AlN structures with random layers' lengths have already
been studied as sources of photon pairs with ultra-narrow spectral
widths that originated in an optical analog of Anderson
localization of the down-converted fields
\cite{PerinaJr2009b,PerinaJr2009c}. It has also been shown that
GaN/AlN nonlinear layers allow the generation of photon-pair
states antisymmetric with respect to the exchange of the signal-
and idler-field frequencies that exhibit anti-bunching and
anti-coalescence \cite{PerinaJr2007b}.

The investigated layered structures as sources of photon pairs can
be compared with other photon-pair sources with respect to
intensity transverse profile, correlated area and efficiency of
photon-pair generation as follows. The vast majority of
photon-pair sources including wave-guiding structures and
nonlinear crystals with/withour poling are designed such that the
signal (and similarly the idler) photon is generated into one
compact and small emission area. In some sources, bulk crystals
are cut in such a way that the spatial symmetry allows to generate
photons around the whole cone surface \cite{Vallone2007}. In this
case the generated state is entangled also in the wave vectors of
the signal and idler photons. As the Hilbert space corresponding
to this degree of freedom has a greater number of independent
states we obtain a higher-dimensional entangled state. In case of
layered structures, the signal (and idler) photon can even be
generated around several concentric cone surfaces depending on
complexity of the structure. This even enlarges the number of
independent states in the transverse area making the generated
state suitable for 'parallel processing' of quantum information in
the transverse plane. As for correlated areas, they are described
by compact profiles (usually of an elliptical shape) in all common
photon-pair sources with cw pumping. Splitting of correlated areas
into several well-separated parts is a distinguished property of
nonlinear layered structures. Even three different mechanisms
leading to this splitting exist in layered structures: zig-zag
movement of photons inside the structure, spatial symmetry and
polarization-dependent properties. The last two mechanisms may in
principle occur also in other photon-pair sources. However,
enhancement of the fields' amplitudes inside a structure is needed
to make them significant. Finally, periodically-poled nonlinear
crystals provide the greatest photon-pair fluxes. On the other
hand, wave-guiding structures including nonlinear planar
wave-guides and nonlinear structured fibers have the greatest
nonlinear conversion efficiencies due to the transverse
confinement of the interacting fields. However, they allow for
only moderate pumping intensities because of a possible material
damage and competing nonlinear processes. Nonlinear layered
structures lie in the middle in this comparison. They provide
greater nonlinear conversion efficiencies than bulk crystals owing
to enhanced electric-field amplitudes along the propagation
direction. Compared to wave-guiding structures, their nonlinear
conversion efficiencies are lower because the 'fields'
confinement' is only in one dimension. On the other hand, similar
pumping intensities as those used for bulk crystals can be
applied.

The paper is organized as follows. A spatial vectorial quantum
model of SPDC in layered media is present in Sec.~II and provides
quantities characterizing the emitted photon pairs. A systematic
method for designing efficient nonlinear structures is described
in Sec.~III. Transverse intensity profiles and correlated areas
are discussed in Sec.~IV using structures with different numbers
of layers. Conclusions are drawn in Sec.~V.

\section{Spatial quantum model of spontaneous parametric
down-conversion}

Nonlinear Hamiltonian $ \hat{H}_{\rm int} $ characterizing SPDC in
a nonlinear medium of volume $ \cal V $ at time $ t $ can be
written in the form \cite{Mandel1995}:
\begin{eqnarray}  
 \hat{H}_{\rm int}(t) &=& \epsilon_0  \int_{\cal V} d{\bf r} \;  \nonumber \\
 & & \hspace{-10mm}  {\bf d}({\bf r}):
 \left[ {\bf E}_{p}^{(+)}({\bf r},t) \hat{\bf E}_{s}^{(-)}({\bf r},t)
 \hat{\bf E}_{i}^{(-)}({\bf r},t) + {\rm h.c.} \right].
\label{1}
\end{eqnarray}
In Eq.~(\ref{1}), $ {\bf d} $ stands for a third-order tensor of
nonlinear coefficients and the symbol $ : $ means shorthand of the
tensor $ {\bf d} $ with respect to its three indices. A strong
pump field is described by the positive-frequency part $ {\bf
E}_{p}^{(+)}({\bf r},t) $ of its electric-field amplitude vector.
A signal [idler] field at single-photon level is characterized by
the negative-frequency part $ \hat{\bf E}_{s}^{(-)}({\bf r},t) $
[$ \hat{\bf E}_{s}^{(-)}({\bf r},t) $] of its electric-field
operator amplitude. Symbol $ \epsilon_0 $ denotes permittivity of
vacuum whereas $ {\rm h.c.} $ stands for the Hermitian conjugate
term.

The positive-frequency amplitudes $ {\bf E}_{m}^{(+)}({\bf r},t) $
of the interacting fields ($ m=p,s,i $) can be in general
decomposed into plane waves with wave vectors $ {\bf k}_m $ and
amplitudes $ {\bf E}_{m}^{(+)}({\bf k}_m) $:
\begin{equation}   
 {\bf E}_{m}^{(+)}({\bf r},t) = \frac{1}{(\sqrt{2\pi})^3} \int d^3{\bf
  k}_m {\bf E}_{m}^{(+)}({\bf k}_m) \exp(i{\bf k}_m{\bf
  r}-i\omega_m t);
\label{2}
\end{equation}
$ \omega_m $ is the frequency of field $ m $ determined in
accordance with dispersion relations.

\begin{figure}    
 \resizebox{0.9\hsize}{!}{\includegraphics{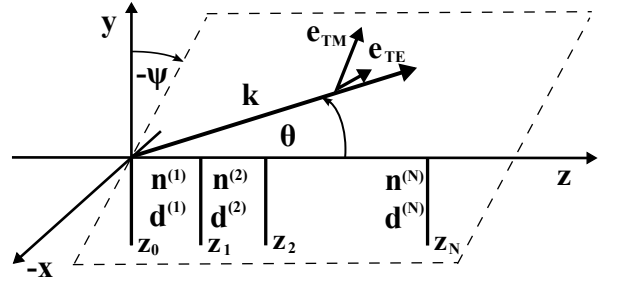}}
 \vspace{2mm}
 \caption{Scheme of the structure and used coordinate system. A plane wave with
  wave vector $ {\bf k} $ propagates along the radial emission angle $ \vartheta $
  and azimuthal emission angle $ \psi $. The radial emission
  angle $ \vartheta $ is measured in the plane of incidence from the $ +z $ axis. The azimuthal
  emission angle $ \psi $ gives rotation in the $ xy $ plane beginning from the $ +y $ axis
  and rotating towards
  the $ -x $ axis. Vectors $ {\bf e}_{\rm TE} $ and $ {\bf e}_{\rm TM} $ give polarization directions
  of TE and TM waves determined with respect to the plane of incidence. Symbols
  $ z_i $ for $ i=0, \ldots, N $ identify $ z $ positions of the boundaries being perpendicular to the
  $ z $ axis; $ n^{(l)} $ [$ {\bf d}^{(l)} $] means index of refraction [tensor of nonlinear coefficients]
  in an $ l $th layer.}
 \label{fig1}
\end{figure}

Considering the incident and un-depleted pump field, we assume
that its temporal spectrum $ {\cal E}_p(\omega_p) $ as well as
spatial spectrum $ {\cal E}_p^{\rm tr}(k_{px},k_{py}) $ in the
transverse plane are given. In this case, the decomposition of
amplitude $ {\bf E}_{p}^{(+)}({\bf r},t) $ in Eq.~(\ref{2}) can be
rewritten as:
\begin{eqnarray}   
 {\bf E}_{p}^{(+)}({\bf r},t) &=& \frac{1}{(\sqrt{2\pi}c)^3} \int_{-\pi/2}^{\pi/2}
  \sin(\vartheta_p) d\vartheta_p \int_{-\pi/2}^{\pi/2} d\psi_p
  \nonumber \\
 & & \hspace{-20mm} \int_{0}^{\infty} \omega_p^2 d\omega_p \; {\cal E}_p(\omega_p)
  {\cal E}_p^{\rm tr} \left[ k_{p,x}
  ({\bf \Omega_p}), k_{p,y}({\bf \Omega_p}) \right]  \nonumber \\
 & & \hspace{-20mm} \mbox{} \times  \exp
  \left[ i k_{p,x}({\bf \Omega_p}) x + i k_{p,y}({\bf \Omega_p}) y\right]
  \sum_{\alpha={\rm TE,TM}} {\bf
  E}_{p,\alpha}^{(+)}(z,{\bf \Omega_p})
  \nonumber \\
 & & \hspace{-20mm} \mbox{} \times  \exp(-i\omega_p t);
\label{3}
\end{eqnarray}
using the vector $ {\bf \Omega}_p \equiv
(\omega_p,\vartheta_p,\psi_p) $ of 'spherical coordinates'  $
\omega_p $, $ \vartheta_p $ and $ \psi_p $. Speed of light in
vacuum is denoted as $ c $. Assuming for simplicity air around the
structure the $ x $ and $ y $ components of wave vector $ {\bf
k}_p $ in front of the structure are given as:
\begin{eqnarray}  
 k_{p,x}({\bf \Omega}_p) &=&
  - \frac{\omega_p\sin(\psi_p)\sin(\vartheta_p) }{c} , \nonumber \\
 k_{p,y}({\bf \Omega}_p) &=& \frac{
  \omega_p \cos(\psi_p)\sin(\vartheta_p) }{c}.
\label{4}
\end{eqnarray}
The decomposition of pump-field amplitude $ {\bf E}_{p}^{(+)} $
into TE and TM waves as given in Eq.~(\ref{3}) is done with
respect to the plane of incidence of the wave with wave vector $
{\bf k}_p $ propagating through the layered structure (see
Fig.~\ref{fig1} for a scheme of the structure). We note that
projections of $ {\bf k} $ vectors into the planes of boundaries
are conserved through the structure. The pump-field amplitudes $
{\bf E}_{p,\alpha}^{(+)} $ introduced in Eq.~(\ref{3}) describe
field evolution along the $ z $ axis where the field undergoes
back-scattering at the boundaries.

Considering a structure with $ N $ layers and boundaries
perpendicular to the $ z $ axis localized at positions $ z_n $, $
n=0, \ldots, N $, the pump-field amplitude $ {\bf
E}_{p,\alpha}^{(+)} $ can be written in the form (for more
details, see \cite{PerinaJr2006}):
\begin{eqnarray}     
 {\bf E}_{p,\alpha}^{(+)}(z,{\bf \Omega}_p) &=&
   {\rm rect}_{-\infty,z_0}(z) \sum_{a=F,B}
   A_{p_a,\alpha}^{(0)}({\bf \Omega}_p) \nonumber \\
  & & \hspace{-1cm} \mbox{} \times {\bf e}_{p_a,\alpha}^{(0)}({\bf \Omega}_p)
  \exp\left[iK_{p_a,z}^{(0)}({\bf \Omega}_p)(z-z_0)\right] \nonumber \\
  & & \mbox{} \hspace{-.3cm} + \sum_{l=1}^{N} {\rm rect}_{z_{l-1},z_l}(z)
   \sum_{a=F,B} A_{p_a,\alpha}^{(l)}({\bf \Omega}_p)\nonumber \\
  & & \mbox{} \hspace{-1cm} \times {\bf e}_{p_a,\alpha}^{(l)}
   ({\bf \Omega}_p) \exp\left[iK_{p_a,z}^{(l)}({\bf \Omega}_p)
   (z-z_{l-1})\right] \nonumber \\
  & & \mbox{} \hspace{-.3cm} + {\rm rect}_{z_N,\infty}(z)
   \sum_{a=F,B} A_{p_a,\alpha}^{(N+1)}({\bf \Omega}_p)\nonumber \\
  & & \mbox{} \hspace{-1cm} \times  {\bf e}_{p_a,\alpha}^{(N+1)}({\bf \Omega}_p)
   \exp\left[iK_{p_a,z}^{(N+1)}({\bf \Omega}_p)(z-z_N) \right]; \nonumber
   \nonumber \\
  & & \alpha={\rm TE,TM}.
\label{5}
\end{eqnarray}
The function $ {\rm rect}_{z_a,z_b}(z) $ equals one for $ z_a \le
z \le z_b $ and is zero otherwise. Symbols $ {\bf
e}^{(l)}_{p_F,\alpha} $ and $ {\bf e}^{(l)}_{p_B,\alpha} $ denote
polarization vectors of $ \alpha $ waves for forward- and
backward-propagating fields with respect to the $ +z $ axis,
respectively.

The $ z $ component $ K_{p_a,z}^{(l)}({\bf \Omega}_p) $ of wave
vector in an $ l $th layer belonging to wave $ a $ with frequency
$ \omega_p $ and propagating in direction ($ \vartheta_p, \psi_p
$) in front of the structure can be expressed as:
\begin{equation}    
 K_{p_a,z}^{(l)}({\bf \Omega}_p) = \pm \frac{ n_p^{(l)}(\omega_p)\omega_p}{c}
 \cos(\vartheta_p^{(l)}),
\label{6}
\end{equation}
where the sign $ + $ ($ - $) is appropriate for a forward-
(backward-) propagating wave. Index of refraction of the pump
field in an $ l $th layer is denoted as $ n_p^{(l)} $. As we
assume that the layered structure is surrounded by air, we have $
n_p^{(0)} = n_p^{(N+1)} = 1 $. The propagation angle $
\vartheta_p^{(l)} $ in the $ l $th layer is then derived from
Snell's law:
\begin{equation}  
 n_p^{(0)}\sin(\vartheta_p^{(0)}) = n_p^{(l)}\sin(\vartheta_p^{(l)}),
  \hspace{5mm} l=1,\ldots,N+1;
\end{equation}
$ \vartheta_p^{(0)} \equiv \vartheta_p $.

The coefficients $ A_{p_F,\alpha}^{(l)}({\bf \Omega}_p) $ and $
A_{p_B,\alpha}^{(l)}({\bf \Omega}_p) $ introduced in Eq.~(\ref{5})
determine amplitudes of $ \alpha $ waves with frequency $ \omega_p
$ propagating forward and backward, respectively, in the ($
\vartheta_p,\psi_p $) direction. These coefficients are derived
from Fresnel's relations at the boundaries using, e.g., the
transfer-matrix formalism \cite{Yeh1988}. Values of the
coefficients $ A_{p_F,\alpha}^{(0)}({\bf \Omega}_p) $ and $
A_{p_B,\alpha}^{(N+1)}({\bf \Omega}_p) $ for $ \alpha = {\rm TE,
TM} $ characterize the pump fields incident on the structure from
both sides and represent the boundary conditions. The
transfer-matrix formalism has been elaborated in detail in
\cite{PerinaJr2006} for layered structures. The coefficients $
A^{(l)} $ of the fields with wave vectors lying in the plane of
incidence given by an angle $ \psi_p $ are given by Eq.~(23) of
Ref.~\cite{PerinaJr2006}.

The signal and idler fields are at single-photon level in the
spontaneous process and so they have to be treated using quantum
theory \cite{Mandel1995}. Their electric-field operator amplitudes
in the layered structure can be decomposed into plane waves using
Eq.~(\ref{1}) and subsequently described in analogy to the pump
field. Using the 'spherical coordinates' $ \omega_m $ $,
\vartheta_m $ and $ \psi_m $ and defining $ {\bf \Omega}_m =
(\omega_m,\vartheta_m,\psi_m) $, $ m=s,i $, we can express the
positive-frequency part $ \hat{E}_{m}^{(+)} $ of the
electric-field operator amplitude of mode $ m $ as:
\begin{eqnarray}   
 \hat{\bf E}_{m}^{(+)}({\bf r},t) &=& \frac{1}{c^3} \int_{-\pi/2}^{\pi/2}
  \sin(\vartheta_m) d\vartheta_m \int_{-\pi/2}^{\pi/2} d\psi_m   \nonumber \\
 & & \hspace{-20mm} \int_0^\infty \omega_m^2 d\omega_m \exp
   \left[i k_{m,x}({\bf \Omega}_m) x +
    i k_{m,y}({\bf \Omega}_m) y\right]
   \nonumber \\
 & & \hspace{-20mm} \mbox{} \times \sum_{\alpha={\rm TE,TM}}
  \sqrt{ \frac{\hbar\omega_m}{16\pi^3\epsilon_0} }
  \; \hat{\bf a}_{m,\alpha}(z,{\bf \Omega}_m) \exp(-i\omega_m t);
\label{8}
\end{eqnarray}
$ \hbar $ stands for the reduced Planck constant and $ \hat{\bf
E}_{m}^{(-)} = \hat{\bf E}_{m}^{(+)\dagger} $ . The expression $
\sqrt{ \hbar\omega_m/(16\pi^3\epsilon_0)} $ gives an
electric-field amplitude per one photon with energy $
\hbar\omega_m $ propagating at speed $ c $. The $ x $ and $ y $
components of wave vectors $ {\bf k}_m $ of the signal and idler
fields ($ m=s,i $) are defined outside the structure:
\begin{eqnarray} 
 k_{m,x}({\bf \Omega}_m) &=&
  - \frac{\omega_m\sin(\psi_m)\sin(\vartheta_m)}{c} , \nonumber \\
 k_{m,y}({\bf \Omega}_m) &=& \frac{
 \omega_m\cos(\psi_m)\sin(\vartheta_m) }{c}.
\label{9}
\end{eqnarray}

The operator amplitudes $ \hat{\bf a}_{m,\alpha}(z,{\bf \Omega}_m)
$ introduced in Eq.~(\ref{8}) can be derived in the considered
layered structure in the form:
\begin{eqnarray}     
 \hat{\bf a}_{m,\alpha}(z,{\bf \Omega}_m) &=&
  {\rm rect}_{-\infty,z_0}(z) \sum_{a=F,B}
   \hat{a}_{m_a,\alpha}^{(0)}({\bf \Omega}_m)
   \nonumber \\
  & & \hspace{-1cm} \mbox{} \times {\bf e}_{m_a,\alpha}^{(0)}({\bf \Omega}_m)
   \exp\left[iK_{m_a,z}^{(0)}
   ({\bf \Omega}_m)(z-z_0)\right] \nonumber \\
  & & \mbox{} \hspace{-0.3cm} + \sum_{l=1}^{N} {\rm rect}_{z_{l-1},z_l}(z)
   \sum_{a=F,B} \hat{a}_{m_a,\alpha}^{(l)}({\bf \Omega}_m)\nonumber \\
  & & \mbox{} \hspace{-1cm} \times {\bf e}_{m_a,\alpha}^{(l)}
   \exp\left[iK_{m_a,z}^{(l)}({\bf \Omega}_m)
   (z-z_{l-1})\right] \nonumber \\
  & & \mbox{} \hspace{-0.3cm} + {\rm rect}_{z_N,\infty}(z)
   \sum_{a=F,B} \hat{a}_{m_a,\alpha}^{(N+1)}({\bf \Omega}_m)\nonumber \\
  & & \mbox{} \hspace{-1cm} \times  {\bf e}_{m_a,\alpha}^{(N+1)}({\bf \Omega}_m)
   \exp\left[iK_{m_a,z}^{(N+1)}
   ({\bf \Omega}_m)(z-z_N) \right];  \nonumber \\
  & & m=s,i; \hspace{5mm} \alpha={\rm TE, TM}.
\label{10}
\end{eqnarray}
Here, the symbols $ {\bf e}^{(l)}_{m_F,\alpha} $ and $ {\bf
e}^{(l)}_{m_B,\alpha} $ stand for polarization vectors of $ \alpha
$ wave of field $ m $ propagating forward and backward,
respectively. The annihilation operators $
\hat{a}_{m_a,\alpha}^{(l)}({\bf \Omega}_m) $ are defined at the
end of an $ l $th layer for the $ \alpha $ wave with frequency $
\omega_m $ of field $ m $ propagating along the direction ($
\vartheta_m, \psi_m $) either forward ($ a=F $) or backward ($ a=B
$).

The $ z $ component $ K_{m_a,z}^{(l)}({\bf \Omega}_m) $ of wave
vector in the $ l $th layer is determined as:
\begin{eqnarray}    
 & & K_{m_a,z}^{(l)}({\bf \Omega}_m) = \pm \frac{ n_m^{(l)}(\omega_m)\omega_m}{c}
 \cos(\vartheta_m^{(l)}), \nonumber \\
 & & \hspace{25mm} m=s,i; \hspace{5mm} a=F,B.
\label{11}
\end{eqnarray}
The sign $ + $ ($ - $) stands for a forward- (backward-)
propagating wave and $ n_m^{(l)} $ gives index of refraction of
field $ m $ in the $ l $th layer. Snell's law provides the
propagation angles $ \vartheta_m^{(l)} $ needed in the $ l $th
layer:
\begin{eqnarray}  
 & & n_m^{(l)}(\omega_m)\sin(\vartheta_m^{(l)}) = n_m^{(N+1)}(\omega_m)\sin(\vartheta_m^{(N+1)}),
  \nonumber \\
 & & \hspace{40mm} \hspace{1mm} l=0,\ldots,N;
\end{eqnarray}
$ \vartheta_m^{(N+1)} \equiv \vartheta_m $.

The operators $ \hat{a}_{m_a,\alpha}^{(l)}({\bf \Omega}_m) $ for $
l=0,\ldots,N+1 $ and fixed $ m $, $ \alpha $, $ \omega_m $, $
\vartheta_m $, and $ \psi_m $ are connected by unitary
transformations at boundaries (Fresnel's relations) and free-field
propagation transformations. This means that the usual boson
commutation relations obeyed by the incident fields are
'transferred' through the structure and the only nonzero
commutation relations \cite{Vogel2001} are the following ones:
\begin{eqnarray}   
 [ \hat{a}_{m_a,\alpha}^{(l)}({\bf \Omega}_m),
   \hat{a}_{{m'}_{a'},\alpha'}^{(l')\dagger}({\bf \Omega}'_{m'})]
   &=& \frac{c^2}{|\sin(\vartheta_m)|\omega_m^2}
   \delta_{m,m'} \delta_{a,a'} \nonumber \\
 & & \hspace{-4.5cm}  \delta_{\alpha,\alpha'} \delta_{l,l'}
   \delta(\omega_m-\omega'_{m'}) \delta(\vartheta_m - \vartheta'_{m'})
   \delta(\psi_m - \psi'_{m'}) .
\label{13}
\end{eqnarray}
The transfer-matrix formalism allows to express the operators $
\hat{a}_{m_a,\alpha}^{(l)}({\bf \Omega}_m) $ in terms of the
operators $ \hat{a}_{m_F,\alpha}^{(N+1)}({\bf \Omega}_m) $ and $
\hat{a}_{m_B,\alpha}^{(0)}({\bf \Omega}_m) $ describing the
outgoing fields. The appropriate relations valid for fields with
wave vectors lying in the plane of incidence given by an angle $
\psi_m $ are given in Eq.~(21) of Ref.~\cite{PerinaJr2006}.

The generation of a photon pair in the process of SPDC governed by
the Hamiltonian $ \hat{H}_{\rm int} $ in Eq.~(\ref{1}) is
described by a perturbation solution of the Schr\"{o}dinger
equation to the first order at time $ t \rightarrow \infty $
assuming an incident vacuum state $ |{\rm vac}\rangle $ in the
signal and idler fields at time $ t \rightarrow -\infty $. The
resulting state $ |\psi\rangle_{s,i}^{\rm out} $ can be derived in
the form:
\begin{eqnarray}    
 |\psi\rangle_{s,i}^{\rm out} &=& |{\rm vac}\rangle - \frac{i}{2c^{8}}
  \sum_{l=1}^{N} \sum_{a,b,g=F,B} \sum_{\alpha, \beta, \gamma={\rm TE,TM}} \nonumber \\
 & & \hspace {-1cm}  \prod_{m=p,s,i} \int_{-\pi/2}^{\pi/2} \sin(\vartheta_m)
  d\vartheta_m \int_{-\pi/2}^{\pi/2} d\psi_m \int_{0}^{\infty} \omega_m^2
  d\omega_m \nonumber \\
 & & \hspace {-1cm} \sqrt{\omega_s\omega_i} \; {\cal E}_p(\omega_p)
  {\cal E}_p^{\rm tr}(k_{p,x},k_{p,y})
   \delta(\omega_p-\omega_s-\omega_i) \nonumber \\
 & &  \hspace{-1cm} \mbox{} \times \delta \left[ k_{p,x} - k_{s,x} - k_{i,x}\right]
  \delta\left[ k_{p,y} - k_{s,y} - k_{i,y}\right] \nonumber \\
 & & \hspace{-1cm} \mbox{} \times  {\bf d}^{(l)}:
  {\bf e}_{p_a,\alpha}^{(l)}({\bf \Omega}_m) {\bf e}_{s_b,\beta}^{(l)*}
  ({\bf \Omega}_s) {\bf e}_{i_g,\gamma}^{(l)*}({\bf \Omega}_i)
  \nonumber \\
 & & \hspace{-1cm} \mbox{} \times \exp\left[\frac{i}{2}
  \Delta K_{p_a s_b i_g,z}^{(l)}({\bf \Omega}_p,{\bf \Omega}_s,{\bf
   \Omega}_i)L_l \right] \nonumber \\
 & & \hspace{-1cm}\mbox{} \times  L_l
  {\rm sinc}\left[\frac{1}{2}\Delta K_{p_a s_b i_g,z}^{(l)}({\bf \Omega}_p,{\bf \Omega}_s,{\bf
   \Omega}_i)L_l \right] \nonumber \\
 & & \hspace{-1cm} \mbox{} \times  A_{p_a,\alpha}^{(l)}({\bf \Omega}_m)
  \hat{a}_{s_b,\beta}^{(l)\dagger}({\bf \Omega}_s)
  \hat{a}_{i_g,\gamma}^{(l)\dagger}({\bf \Omega}_i)
  |{\rm vac} \rangle.
  \label{14}
\end{eqnarray}
The functions $ \Delta K_{p_a s_b i_g,z}^{(l)}({\bf \Omega}_p,{\bf
\Omega}_s,{\bf \Omega}_i) = K_{p_a,z}^{(l)}({\bf \Omega}_p) -
K_{s_b,z}^{(l)}({\bf \Omega}_s) - K_{i_g,z}^{(l)}({\bf \Omega}_i)
$ stand for phase mismatches in the $ l $th layer. Symbol $ L_l $
means the length of $ l $th layer ($ L_l = z_l - z_{l-1} $). The
transverse wave vectors $ k_{m,x}({\bf \Omega}_m) $ and $
k_{m,y}({\bf \Omega}_m) $ for $ m=p,s,i $ are defined in
Eqs.~(\ref{4}) and (\ref{9}) and characterize the fields outside
the structure. We note that the approach based on the solution of
Schr\"{o}dinger equation does not take into account surface SPDC
that generates additional photon pairs
\cite{PerinaJr2009d,PerinaJr2009e}.

The phase-matching conditions in the transverse plane $ xy $ are
described by two $ \delta $ functions in Eq.~(\ref{14}), that
determine the emission direction ($ \vartheta_i,\psi_i $) of an
idler photon provided that the signal-photon emission direction ($
\vartheta_s,\psi_s $) is given and the pump-field is in the form
of a plane wave propagating in direction ($ \vartheta_p,\psi_p $).
Simple geometric considerations provide the following formulas:
\begin{eqnarray}   
 \psi_i &=& \psi_p + \arctan\left[\frac{ \omega_s\sin(\vartheta_s)\sin(\psi_p-\psi_s) }{ \omega_p
   \sin(\vartheta_p) -
   \omega_s\sin(\vartheta_s)\cos(\psi_p-\psi_s)} \right] , \nonumber \\
 \vartheta_i &=& \arcsin\left[ \frac{ \omega_p\sin(\vartheta_p)
  }{ \omega_i\cos(\psi_p-\psi_i) } \right. \nonumber \\
 & & \mbox{} \hspace{2cm} \left. - \frac{ \omega_s
  \cos(\psi_p-\psi_s) }{ \omega_i\cos(\psi_p-\psi_i) }
  \sin(\vartheta_s) \right] .
\label{15}
\end{eqnarray}
If the pump beam is focused, the strict phase-matching conditions
in the transverse plane formulated in Eq.~(\ref{15}) are relaxed
and we arrive at correlation areas with finite spreads.

The expression in Eq.~(\ref{14}) for the state $
|\psi\rangle_{si}^{\rm out} $ can be rewritten into the form
containing only the outgoing creation operators $
\hat{a}_{m_F,\beta}^{(N+1)\dagger}({\bf \Omega}_m) $ and $
\hat{a}_{m_B,\beta}^{(0)\dagger}({\bf \Omega}_m) $ for $ m=s,i $
and $ \beta = {\rm TE, TM} $ using the formulas in Eqs.~(21) and
(23) of Ref.~\cite{PerinaJr2006}. The outgoing operators can be
finally transformed into the polarization basis of the detector
using a suitable unitary transformation. We assume that the
detection plane is perpendicular to the field propagation
direction ($ \vartheta, \psi $) and its $ s $-polarization
(denoted as $ \perp $) is parallel to the horizontal plane $ xz $;
$ p $-polarization (denoted as $ \| $) is defined by the
orthogonality conditions. Assuming field $ m $ ($ m=s,i $) at
frequency $ \omega_m $ and propagating along the angles $
\vartheta_m $ and $ \psi_m $ the needed $ \vartheta_m $- and $
\psi_m $-dependent unitary transformation can be written in the
form:
\begin{eqnarray}   
 \left[ \matrix{ \hat{a}_{m_F,{\rm TE}}^{(N+1)}({\bf \Omega}_m) \cr \hat{a}_{m_F,{\rm TM}}^{(N+1)}
  ({\bf \Omega}_m) } \right] &=&  \nonumber \\
  & & \hspace{-1cm} \left[ \matrix{ \cos(\zeta_m) & \sin(\zeta_m)
    \cr -\sin(\zeta_m) & \cos(\zeta_m) } \right] \left[ \matrix{ \hat{a}_{m_F,\perp}({\bf \Omega}_m) \cr
  \hat{a}_{m_F,\|}({\bf \Omega}_m) } \right],
  \nonumber \\
 \left[ \matrix{ \hat{a}_{m_B,{\rm TE}}^{(0)}({\bf \Omega}_m) \cr
  \hat{a}_{m_B,{\rm TM}}^{(0)}({\bf \Omega}_m) } \right] &=& \nonumber \\
  & & \hspace{-1cm}
   \left[ \matrix{ \cos(\zeta_m) & \sin(\zeta_m)
    \cr -\sin(\zeta_m) & \cos(\zeta_m) } \right] \left[
   \matrix{ \hat{a}_{m_B,\perp}({\bf \Omega}_m) \cr
  \hat{a}_{m_B,\|}({\bf \Omega}_m) } \right],
  \nonumber \\
 \zeta_m(\vartheta_m,\psi_m) &=& \arccos\left[
  \frac{ \cos(\psi_m) }{ \sqrt{1+\sin^2(\psi_m)\tan^2(\theta_m)} }
  \right]  \nonumber \\
 & & \mbox{} \times {\rm sign}(\psi_m) , \hspace{10mm} m=s,i .
\label{16}
\end{eqnarray}
The function $ {\rm sign} $ gives the sign of its argument. The
newly introduced annihilation operators $
\hat{a}_{m_b,\alpha}({\bf \Omega}_m) $, $ m=s,i $, $ b=F,B $, $
\alpha=\|,\perp $, describe the signal and idler fields in the
polarization bases connected with the detectors.

The terms of state $ |\psi\rangle^{\rm out}_{s,i} $ in
Eq.~(\ref{14}) describing the created photon pair can be
decomposed into four groups according to the propagation
directions of the signal and idler photons with respect to the $
+z $ axis ($ FF $, $ FB $, $ BF $, $ BB $). Inside these groups
there occur four contributions that differ in signal- and
idler-photon polarization directions ($ \| \, \| $, $ \| \, \perp
$, $ \perp \, \| $, $ \perp \, \perp $). Each contribution can be
written in the form:
\begin{eqnarray}   
 |\psi\rangle_{s_a,i_b}^{\alpha,\beta}({\bf r},t) &=&
  \prod_{m=s,i} \int_{-\pi/2}^{\pi/2} \sin(\vartheta_m)
   d\vartheta_m \int_{-\pi/2}^{\pi/2} d\psi_m  \nonumber \\
 & & \hspace{-2.3cm}  \int_{0}^{\infty}
   d\omega_m \phi_{ab}^{\alpha,\beta}
   ({\bf \Omega}_s,{\bf \Omega}_i)
     \hat{a}_{s_a,\alpha}^\dagger({\bf \Omega}_s)
     \hat{a}_{i_b,\beta}^\dagger({\bf \Omega}_i) |{\rm vac} \rangle  \nonumber \\
 & & \mbox{} \hspace{-2.3cm} \times
  \exp[-i({\bf k}_{s_a}^{\rm out} + {\bf k}_{i_b}^{\rm out}){\bf r}]
  \exp[i(\omega_s+\omega_i) t]; \nonumber \\
 & & \hspace{0.5cm} a,b = F,B; \hspace{5mm}\alpha,\beta = \|, \perp .
 \label{17}
\end{eqnarray}
The wave vectors $ {\bf k}_{s_a}^{\rm out} $ and $ {\bf
k}_{i_b}^{\rm out} $ characterize free-field evolution of the
emitted signal and idler fields, respectively, outside the
structure. The functions $ \phi_{ab}^{\alpha,\beta} ({\bf
\Omega}_s,{\bf \Omega}_i) $ introduced in Eq.~(\ref{17})
characterize completely properties of the generated photon pair.
It gives the probability amplitude of having an $ \alpha
$-polarized signal photon at frequency $ \omega_s $ propagating
along direction ($ \vartheta_s, \psi_s $) together with its $
\beta $-polarized idler photon at frequency $ \omega_i $
propagating along direction ($ \vartheta_i, \psi_i $) at the
output $ ab $ of the structure.

Intensity spatial and spectral properties of a photon pair
\cite{PerinaJr2006} can be conveniently derived from a density $
n_{ab}^{\alpha,\beta} $ of the mean photon-pair numbers belonging
to the state $ |\psi\rangle_{s_a,i_b}^{\alpha,\beta} $. The
density $ n_{ab}^{\alpha,\beta} $ is defined along the expression
\begin{eqnarray}       
 n_{ab}^{\alpha,\beta}({\bf \Omega}_s,{\bf \Omega}_i) =
 {}_{s_a,i_b}^{\alpha,\beta}\langle \psi| \hat{n}_{s_a,\alpha}({\bf \Omega}_s)
  \hat{n}_{i_b,\beta}({\bf \Omega}_i)
  |\psi\rangle_{s_a,i_b}^{\alpha,\beta}.
\label{18}
\end{eqnarray}
The photon-number density operator $ \hat{n}_{m_a,\alpha}({\bf
\Omega}_m) $ is expressed as
\begin{eqnarray}       
 \hat{n}_{m_a,\alpha}({\bf \Omega}_m) &=& \hat{a}_{m_a,\alpha}^\dagger
  ({\bf \Omega}_m) \hat{a}_{m_a,\alpha}({\bf \Omega}_m).
\end{eqnarray}
Using Eq. (\ref{17}) the formula for density $
n_{ab}^{\alpha,\beta} $ of mean photon-pair numbers in Eq.
(\ref{18}) attains a simple form:
\begin{equation}    
 n_{ab}^{\alpha,\beta}({\bf \Omega}_s,{\bf \Omega}_i) =
  |\phi_{ab}^{\alpha,\beta}({\bf \Omega}_s,{\bf \Omega}_i)|^2.
\label{20}
\end{equation}

A density $ n_{s,ab}^{\alpha,\beta}({\bf \Omega}_s) $ of mean
signal-photon numbers in the state $
|\psi\rangle_{s_a,i_b}^{\alpha,\beta} $ can easily be determined
from the density $ n_{ab}^{\alpha,\beta} $ defined in
Eq.~(\ref{18}) and using the relation in Eq.~(\ref{20}):
\begin{eqnarray}   
 n_{s,ab}^{\alpha,\beta}({\bf \Omega}_s) &=& \int_{-\pi/2}^{\pi/2}
  \sin(\vartheta_i) d\vartheta_i  \int_{-\pi/2}^{\pi/2} d\psi_i
  \int_{0}^{\infty} d\omega_i \nonumber \\
 & &  |\phi_{ab}^{\alpha,\beta}({\bf \Omega}_s,{\bf \Omega}_i)|^2.
\label{21}
\end{eqnarray}

If spectral resolution in detection of the signal-field transverse
profile is not available, a spatial density $ n_{s,ab}^{{\rm
tr},\alpha,\beta}(\vartheta_s,\psi_s) $ of mean signal-photon
numbers emitted in the direction ($ \vartheta_s,\psi_s $) is a
useful characteristic. It can be derived using the density $
n_{s,ab}^{\alpha,\beta} $ of mean signal-photon numbers given in
Eq.~(\ref{21}):
\begin{eqnarray}   
 n_{s,ab}^{{\rm tr},\alpha,\beta}(\vartheta_s,\psi_s) &=& \int_{0}^{\infty}
  d\omega_s n_{s,ab}^{\alpha,\beta}({\bf \Omega}_s).
\label{22}
\end{eqnarray}
The above densities can be analogously defined also for the idler
field.

Spatial correlations between the signal- and idler-field photon
numbers in their transverse planes can be quantified in terms of
the fourth-order correlation functions $ n_{ab}^{{\rm
cor},\alpha,\beta}(\vartheta_s,\psi_s,\vartheta_i,\psi_i ) $ that
give joint densities of photon-pair numbers such that a signal
photon propagates along the ($ \vartheta_s,\psi_s $) direction and
the idler twin propagates along the ($ \vartheta_i,\psi_i $)
direction:
\begin{eqnarray}   
  n_{ab}^{{\rm cor},\alpha,\beta}(\vartheta_s,\psi_s,\vartheta_i,\psi_i)
   &=& \int_{0}^{\infty} d\omega_s \int_{0}^{\infty} d\omega_i
   n_{ab}^{\alpha,\beta}({\bf \Omega}_s,{\bf \Omega}_i).   \nonumber \\
   & &
\label{23}
\end{eqnarray}
Provided that the signal-photon propagation direction ($
\vartheta_s^0,\psi_s^0 $) is given, the joint density $
n_{ab}^{{\rm
cor},\alpha,\beta}(\vartheta_s^0,\psi_s^0,\vartheta_i,\psi_i )$
remains a function of the idler-field emission angles $
\vartheta_i $ and $ \psi_i $ and its profile defines a correlated
area. The correlated area of an idler photon determines an area in
the idler-field transverse plane where an idler photon can be
expected provided that its signal twin has been detected in the ($
\vartheta_s^0,\psi_s^0 $) direction.

Finally, an overall mean photon-pair number $
N_{ab}^{\alpha,\beta} $ related to the state $
|\psi\rangle_{s_a,i_b}^{\alpha,\beta} $ can be found using the
following relation:
\begin{eqnarray}    
 N_{ab}^{\alpha,\beta} &=& \prod_{m=s,i} \int_{-\pi/2}^{\pi/2} \sin(\vartheta_m)
   d\vartheta_m \int_{-\pi/2}^{\pi/2} d\psi_m \int_{0}^{\infty}
   d\omega_m  \nonumber \\
 & & \hspace{5mm} n_{ab}^{\alpha,\beta}({\bf \Omega}_s,{\bf \Omega}_i).
\label{24}
\end{eqnarray}

An important feature of layered structures is an increase of the
efficiency of nonlinear process due to the enhanced electric-field
amplitudes caused by interference along the $ z $ axis. This
increase can be quantified with respect to a certain reference
structure which fully exploits the nonlinearity, but does not rely
on interference. In this reference structure, there occurs no
back-scattering of the propagating fields and also the nonlinear
process is assumed to be fully phase matched. The orientations of
nonlinear layers and polarizations of the interacting fields are
such that the most intense nonlinear effect occurs. This reference
structure provides an ideal photon pair with a signal photon
emitted in an arbitrary direction ($ \vartheta_s,\psi_s $) and an
idler photon in the corresponding ($ \vartheta_i,\psi_i $)
direction. Fixing $ \vartheta_s $ and $ \psi_s $, the emitted pair
can be described by the following output state $
|\psi\rangle_{s,i}^{\rm ref} $ [compare Eq.~(\ref{14})]:
\begin{eqnarray}    
 |\psi\rangle_{s,i}^{\rm ref} &=& - \frac{i}{2c^8}
  \int_{0}^{\infty} \, \omega_s^2 d\omega_s \int_{0}^{\infty} \, \omega_i^2
  d\omega_i \sqrt{\omega_s \omega_i} \nonumber \\
 & & \hspace{-2cm} \mbox{} {\cal E}_{p}(\omega_s+\omega_i) \sum_{l=1}^{N}
  \max({\bf d}^{(l)}) L_l
  \hat{a}_{s}^\dagger(\omega_s) \hat{a}_{i}^\dagger(\omega_i)
  |{\rm vac} \rangle,
\label{25}
\end{eqnarray}
where $ \hat{a}_{s}^{\dagger}(\omega_s) $ [$
\hat{a}_{i}^{\dagger}(\omega_i) $] stands for a signal- [idler-]
field creation operator outside the reference structure. The
function $ \max $ used in Eq.~(\ref{25}) gives the maximum value
among the elements of tensor $ {\bf d}^{(l)} $.

Using the reference structure a relative density $
\eta_{s,ab}^{\alpha,\beta}({\bf \Omega}_s) $ of mean signal-photon
numbers belonging to the state $
|\psi\rangle_{s_a,i_b}^{\alpha,\beta} $ can naturally be defined
as
\begin{equation}   
 \eta_{s,ab}^{\alpha,\beta}({\bf \Omega}_s) = \frac{
  n_{s,ab}^{\alpha,\beta}({\bf \Omega}_s) }{ n_s^{\rm ref}(\omega_s) } ,
\label{26}
\end{equation}
where the signal-field photon-number density $
n_{s,ab}^{\alpha,\beta} $ is written in Eq.~(\ref{21}). The
reference signal-field photon-number density $ n_s^{\rm ref} $
characterizes the state $ |\psi\rangle_{s,i}^{\rm ref} $ in
Eq.~(\ref{25}) and does not depend on the propagation angles $
\vartheta_s $ and $ \psi_s $.

Properties of photon pairs in the time domain are complementary to
those found in the spectral domain and belong to important
characteristics of photon pairs. We can mention signal- and
idler-field photon fluxes or coincidence-count interference
patterns in different kinds of interferometers as examples. These
photon-pair properties can be investigated, e.g., using the
formulas contained in Sec.~IIC of Ref. \cite{PerinaJr2006} even in
this spatial vectorial model.

Numerical calculations that follow are performed for a cw pump
field with a Gaussian transverse profile described as follows:
\begin{eqnarray}   
 {\cal E}_p(\omega_p) &=& \xi_p \delta(\omega_p - \omega_p^0 ) ,
  \nonumber \\
 {\cal E}_p^{\rm tr}(k_x,k_y) &=& \frac{r_p}{\sqrt{2\pi}}
  \exp\left[ \frac{ r^2_p(k_x^2 + k_y^2)}{4} \right] .
\label{27}
\end{eqnarray}
In Eq.~(\ref{27}), $ \xi_p $ gives an amplitude of the pump field
with the carrying frequency $ \omega_p^0 $; $ r_p $ denotes the
width of amplitude transverse profile. Normalization of the
function $ {\cal E}_p^{\rm tr} $ is such that $ \int dk_x \int
dk_y |{\cal E}_p^{\rm tr}(k_x,k_y)|^2 = 1 $. In case of cw
pumping, there occur formulas in the above equations that contain
a formal expression $ \delta^2(\omega) $. This expression has to
be replaced by the expression $ 2T/(2\pi) \delta(\omega) $, where
the detection interval extends over $ (-T,T) $ and the limit $ T
\rightarrow \infty $ may be considered.

\section{Design of an efficient layered structure}

Here we consider structures with odd numbers $ N $ of layers made
of two kinds of materials. Layers of material $ b $ of length $
l_b $ are sandwiched by layers of material $ a $ having lengths $
l_a $. Also pumping at a defined carrying frequency $ \omega_p^0 $
and impinging on the structure at normal incidence is assumed. The
down-converted signal and idler fields are assumed to have nearly
degenerate frequencies.

We suggest a method for designing an efficient layered structure
from the point of view of, in general, three-mode nonlinear
interaction. It is based on two observations:
\begin{itemize}
 \item An efficient nonlinear process occurs provided that all three
  nonlinearly interacting fields lie inside their transmission
  peaks. This behavior originates in the fact that the
  electric-field amplitudes of monochromatic fields with frequencies
  in transmission peaks are enhanced inside the structure owing to
  constructive interference of back-scattered light. It follows from
  the band-gap theory that the closer the transmission peak to a
  band gap, the greater the electric-field amplitudes.
 \item As numerical calculations have revealed, the overlap integral over the
  amplitudes of three interacting fields giving the strength of the
  effective nonlinear interaction [see Eq.~(\ref{1})] vanishes if
  the signal- and idler-field amplitudes along the structure are the
  same. This means that photon-pair states degenerate in
  frequencies, emitted in symmetric directions and having the same
  polarizations cannot be generated.
\end{itemize}
We note that these facts have been found crucial in designing
layered structures efficient for second-harmonic generation
\cite{Scalora1997}. Considering a collinear interaction, the
requirements are even more strict because of only one propagation
direction. This requires specific approaches that rely, e.g., on
tuning the frequencies of transmission peaks by changing the index
of refraction of one type of the layers \cite{Scalora1997}.
Alternatively, non-collinear second-harmonic generation has been
considered. In this case, there exists one free parameter (radial
emission angle) that can be varied in order to find an efficient
structure. This represents an equivalent problem to that
considered here and can be treated by the developed systematic
approach.

Returning back to the considered layered structures they are
characterized by three parameters: number $ N $ of layers and
lengths $ l_a $ and $ l_b $ of these layers. A detailed inspection
has revealed that the number $ N $ of layers significantly
determines the number of generated photon pairs as well as angular
extensions of the emission areas of photons in a pair (see
Fig.~\ref{fig3} bellow). The greater the number $ N $ of layers
the greater the number of generated photon pairs and also the
smaller the angular extensions of emission areas. From practical
point of view, the number $ N $ of layers is approximately fixed
considering these dependencies. Having the number $ N $ of layers
fixed, there remain two adjustable parameters - layers' lengths $
l_a $ and $ l_b $. However, these two lengths cannot be chosen
arbitrarily because the pump field at the carrying frequency $
\omega_p^0 $ has to be in a transmission peak.

Let us fix the lengths $ l_a $ and $ l_b $ (together with the
number $ N $ of layers) for a moment and determine the spectral
intensity transmission $ T(\omega) $ along the $ + z $ direction
(of the pump-field propagation) using, e.g., the transfer-matrix
formalism \cite{Yeh1988}. Increasing the frequency $ \omega $
there occur forbidden bands one following the other. The
difference in central frequencies of the neighbor forbidden bands
is roughly the same in accord with the band-gap theory. This may
be convenient for spectrally nearly degenerate SPDC provided that
the process of SPDC can be tuned such that the pump-field
frequency lies inside a transmission peak close to the second
forbidden band whereas the signal- and idler-field frequencies are
inside the transmission peaks near to the first forbidden band. We
note that all three interacting fields can accommodate themselves
either to the transmission peaks above or bellow the forbidden
bands. This helps to obey the quite strong requirements of the
three-field interaction by varying the signal-field emission angle
$ \vartheta_s $. However, suitable conditions can only be revealed
numerically.

Now we return back to the problem in which the pump-field carrying
frequency $ \omega_p^0 $ is fixed. As follows from the
considerations of the previous paragraph, useful structures are
those that have the first upper or the first lower transmission
peak near to the second forbidden band at the pump-field carrying
frequency $ \omega_p^0 $. Inspection of the behavior of structures
with different layers' lengths $ l_a $ and $ l_b $ has shown that
there exists a system of curves in the plane spanned by layers'
lengths $ l_a $ and $ l_b $ that provides the required
transmission peaks at the frequency $ \omega_p^0 $. One curve
corresponds to the first lower transmission peak of the second
forbidden band. Similarly, another curve is associated with the
first upper transmission peak of the second forbidden band. They
can be revealed as follows using the scaling property of
diffraction phenomena in optics.

We consider a fictitious dispersion-free structure with the
indexes of refraction appropriate for the pump-field frequency $
\omega_p^0 $ and define the corresponding optical lengths $
l_a^{\rm opt} $ and $ l_b^{\rm opt} $. It can be shown that the
ratio $ L = l_b^{{\rm opt}}/l_a^{{\rm opt}} $ of optical lengths
represents a suitable variable for parametrization of these
curves. Suitable optical lengths $ l_a^{\rm opt} $ and $ l_b^{\rm
opt} $ for a given value of the ratio $ L $ can be revealed easily
using the scaling property of diffraction phenomena. We fix the
value of length $ l_a^{\rm opt} $ to, e.g., $ l_a^{{\rm opt},0} =
\lambda_p^0/2 = \pi c/\omega_p^0 $. This gives the unit length of
diffraction phenomena that can be, in principle, chosen
arbitrarily. As the ratio $ L $ is given, the optical length $
l_b^{{\rm opt},0} $ is derived as $ l_b^{{\rm opt},0} = L
l_a^{{\rm opt},0} $ and we can calculate the spectral intensity
transmission $ T_p(\omega_p) $ for this structure. We further
identify the frequency $ \omega_p^{\rm max} $ of the first lower
(or upper) transmission peak of the second forbidden band. Then we
have to 'transfer' the actual frequency $ \omega_p^{\rm max} $ of
the transmission peak to the required frequency $ \omega_p^0 $
using the scaling property. The scaling property provides the
layers' optical lengths in the form: $ l_a^{\rm opt} = l_a^{{\rm
opt},0} \omega_p^0 / \omega_p^{\rm max} $ and $ l_b^{\rm opt} = L
l_a^{\rm opt} $. We note that the obtained lengths differ for the
lower and the upper transmission peaks.

In the next step we move along the obtained two curves (for the
upper and the lower transmission peaks) parameterized by the ratio
$ L $ and numerically analyze the structures. The maximum $
\eta_s^{\rm max} $ of relative density $
\eta_s(\omega_s,\vartheta_s, \psi_s^0) $ of mean signal-photon
numbers taken over the signal-field frequency $ \omega_s $ and
radial emission angle $ \vartheta_s $ assuming the fixed
signal-field azimuthal emission angle $ \psi_s^0 $ has been found
a suitable quantity for monitoring efficiency of the nonlinear
process. The signal- and idler-field polarizations are assumed to
be fixed. The greater the value of maximum $ \eta_s^{\rm max} $ of
relative density the closer the signal- and idler-field
transmission peaks to the first forbidden band. A curve giving the
dependence of maximum $ \eta_s^{\rm max} $ of relative density on
the ratio $ L $ is thus a good indicator for choosing suitable
layers' lengths. We note that the curve depends on polarization
properties of the signal and idler fields as well as the azimuthal
signal-field emission angle $ \psi_s^0 $. It is also possible to
monitor another quantity in this procedure, e.g., the overall mean
photon-pair number $ N $ given in Eq.~(\ref{24}). Or the density
of mean signal-photon numbers emitted for a fixed azimuthal
emission angle $ \psi_s^0 $ and determined as $
\int_{-\pi/2}^{\pi/2} \sin(\vartheta_s) d\vartheta_s n_s^{\rm
tr}(\vartheta_s,\psi_s) $ using the formula in Eq.~(\ref{22}) may
be considered.

As an example, the dependence of maximum $ \eta_s^{\perp,\|,\rm
max} $ of relative density of two photons propagating along the $
+z $ axis on the ratio $ L $ for the first lower and the first
upper pump-field transmission peak of the second forbidden band of
the structure composed of 11 and 101 layers, respectively, is
plotted in Fig.~\ref{fig2} (for details, see below). The curve in
Fig.~\ref{fig2}(a) appropriate for the structure with 11 layers is
continuous and rather flat. This means that there exists a whole
continuous set of structures giving an efficient nonlinear
process. On the other hand, efficient nonlinear structures with $
N=101 $ layers are found only in peaks of the curve in
Fig.~\ref{fig2}(b). This is a consequence of complex interference
of back-scattered light inside these structures. We note that the
appropriate transmission peaks in all three interacting fields do
not necessarily exist for all values of the ratio $ L $. Also the
greatest values of maximum $ \eta_s^{\perp,\|,\rm max} $ of the
relative density are found for the values of ratio $ L $ in a
certain restricted region. As the graph in Fig.~\ref{fig2}(b)
shows, the greatest values of maximum $ \eta_s^{\perp,\|,\rm max}
$ occur around $ L = 0.5 $.
\begin{figure}    
 (a) \resizebox{0.9\hsize}{!}{\includegraphics{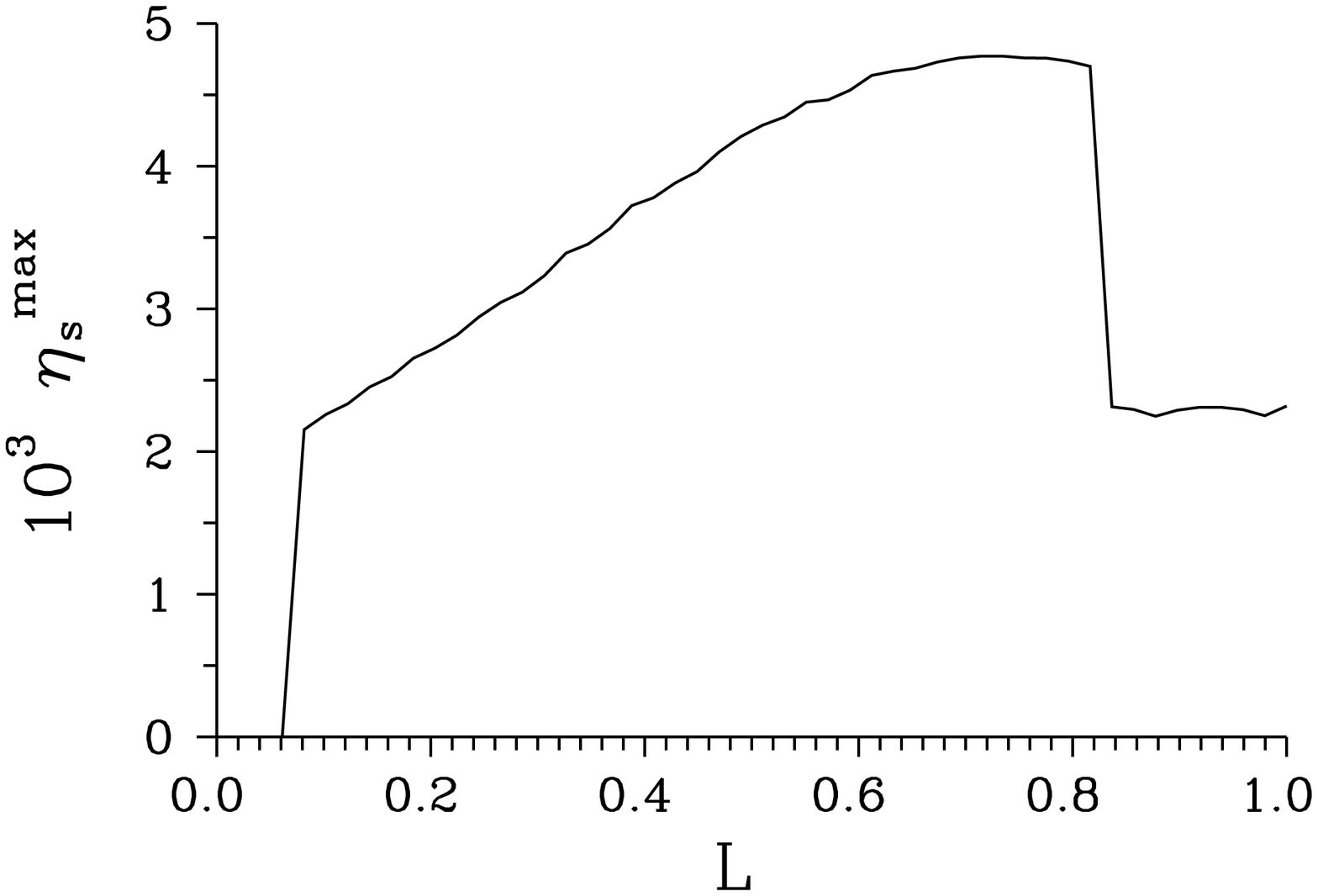}}

 \vspace{2mm}

 (b) \resizebox{0.9\hsize}{!}{\includegraphics{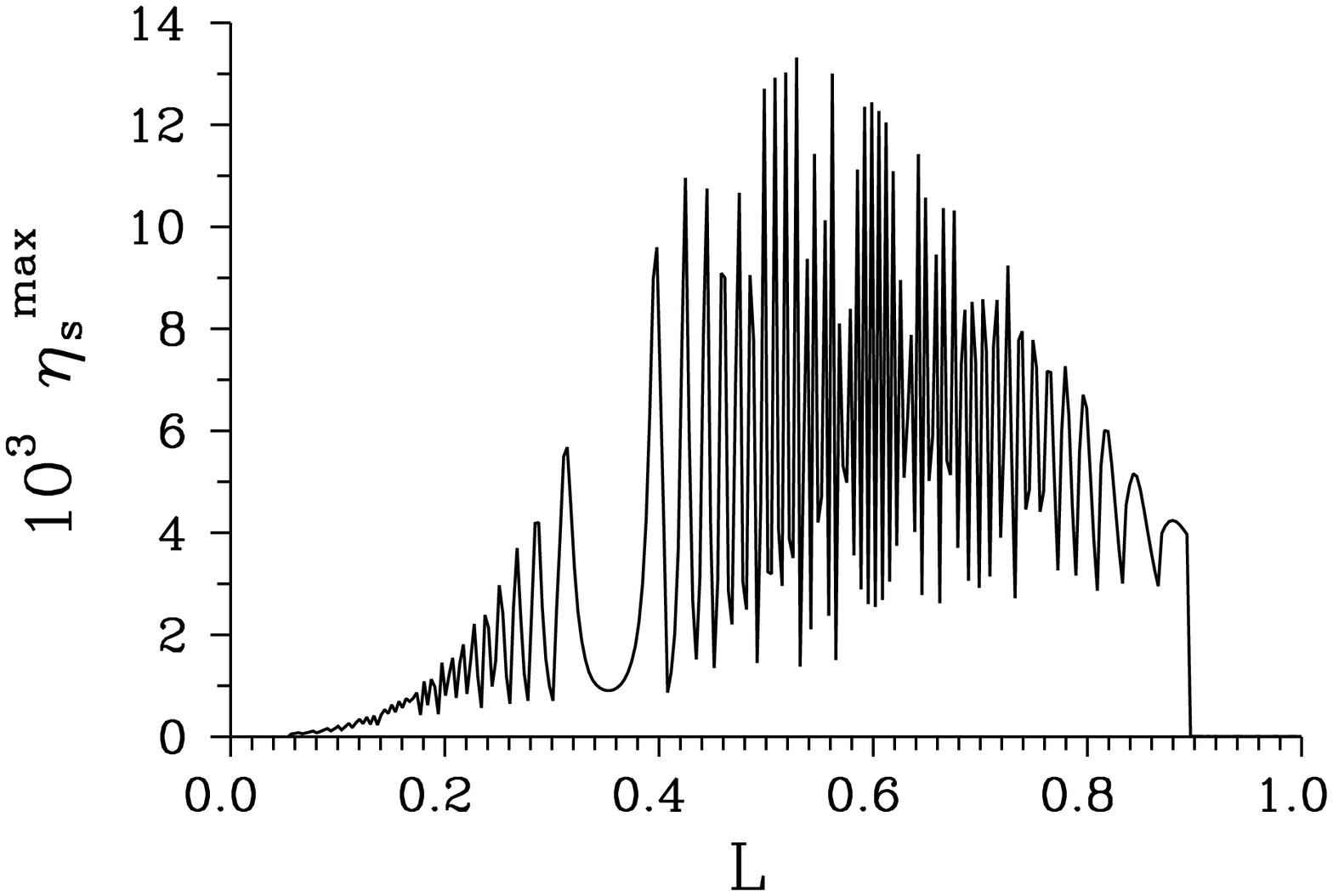}}

 \caption{Maximum $ \eta_s^{\perp,\|,\rm max} $ of the relative
 density of mean signal-photon numbers for states with photons propagating
 forward as a function of the ratio $ L $ of optical layers' lengths for
 the first lower [upper] pump-field
 transmission peak of the second forbidden band for the structure with
 $ N=11 $ [$ N=101 $] layers; $ \psi_s^0 = 0 $~deg.}
\label{fig2}
\end{figure}

The curves in Fig.~\ref{fig2} demonstrate a 'weak' increase of
values of the maximum $ \eta_s^{\perp,\|,\rm max} $ of relative
signal-field density with the increasing values of the number $ N
$ of layers. As the densities $ n_s^{\rm ref} $ of the mean
signal-photon numbers of the reference structure are linearly
proportional to the second power of the length of nonlinear
material inside the structure, the densities $ n_s $ of the mean
signal-photon numbers increase better than $ N^2 $. This is very
important, because the mean number of emitted photon pairs
increases less than the second power of the length of nonlinear
material in other structures producing photon pairs like bulk
crystals, periodically-poled crystals or wave-guiding structures
because of 'longitudinal' phase-matching conditions. Nearly linear
dependence of the mean photon-pair numbers on the length of a
nonlinear structure is commonly observed.

\section{Intensity transverse profiles and correlated areas}

\begin{figure}    
 (a) \resizebox{0.8\hsize}{!}{\includegraphics{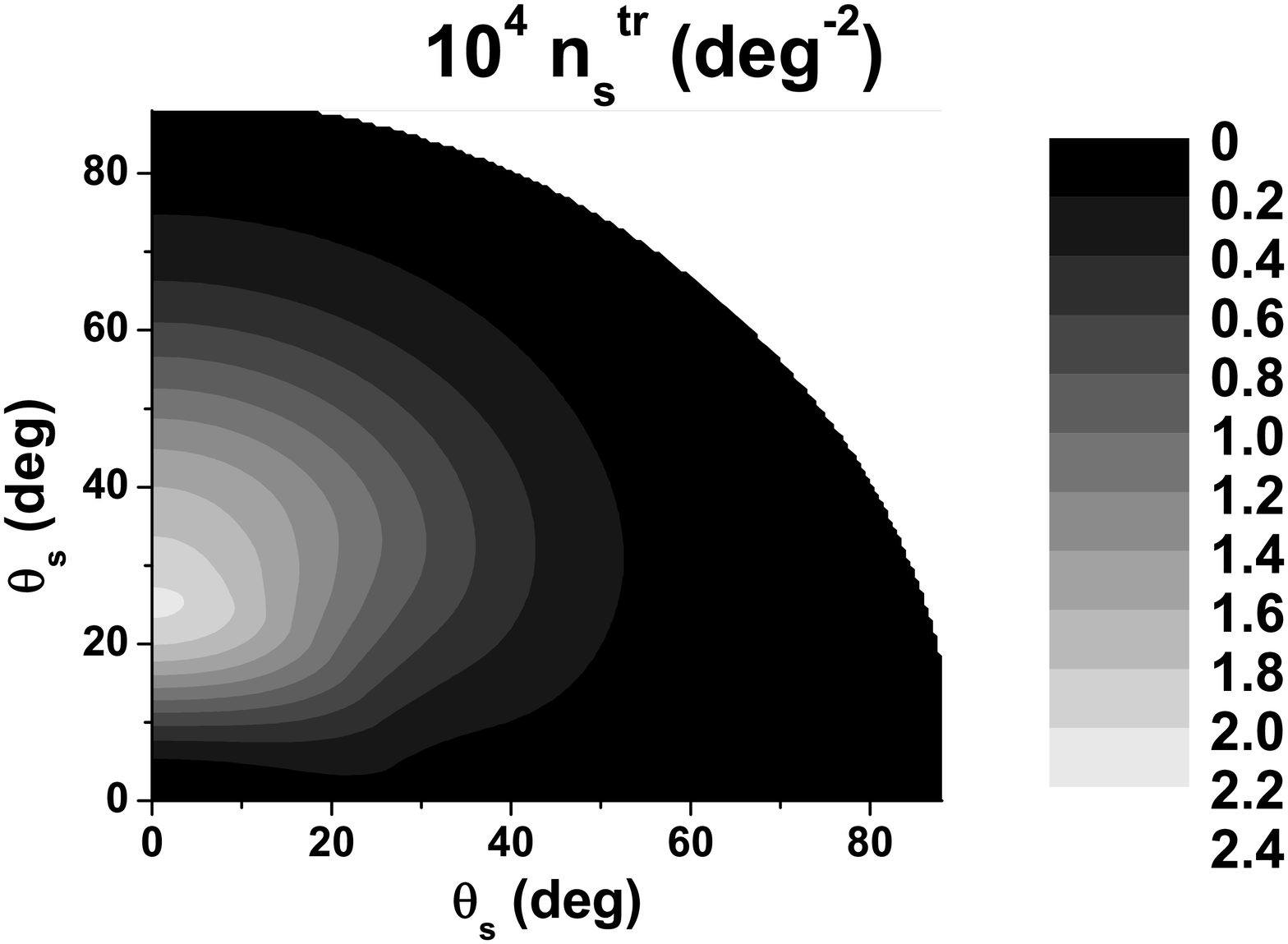}}

 \vspace{2mm}

 (b) \resizebox{0.8\hsize}{!}{\includegraphics{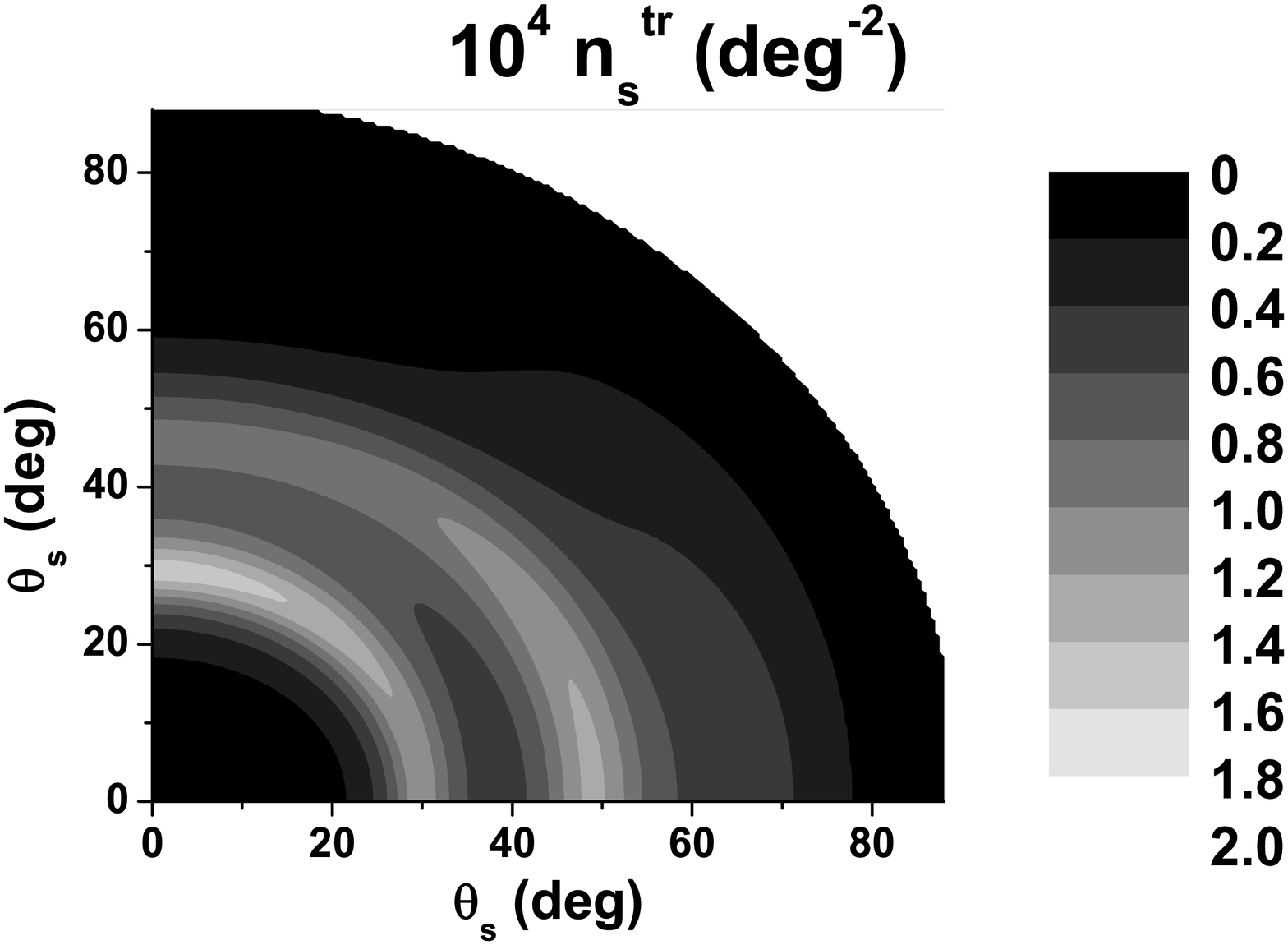}}

 \vspace{2mm}

 (c) \resizebox{0.8\hsize}{!}{\includegraphics{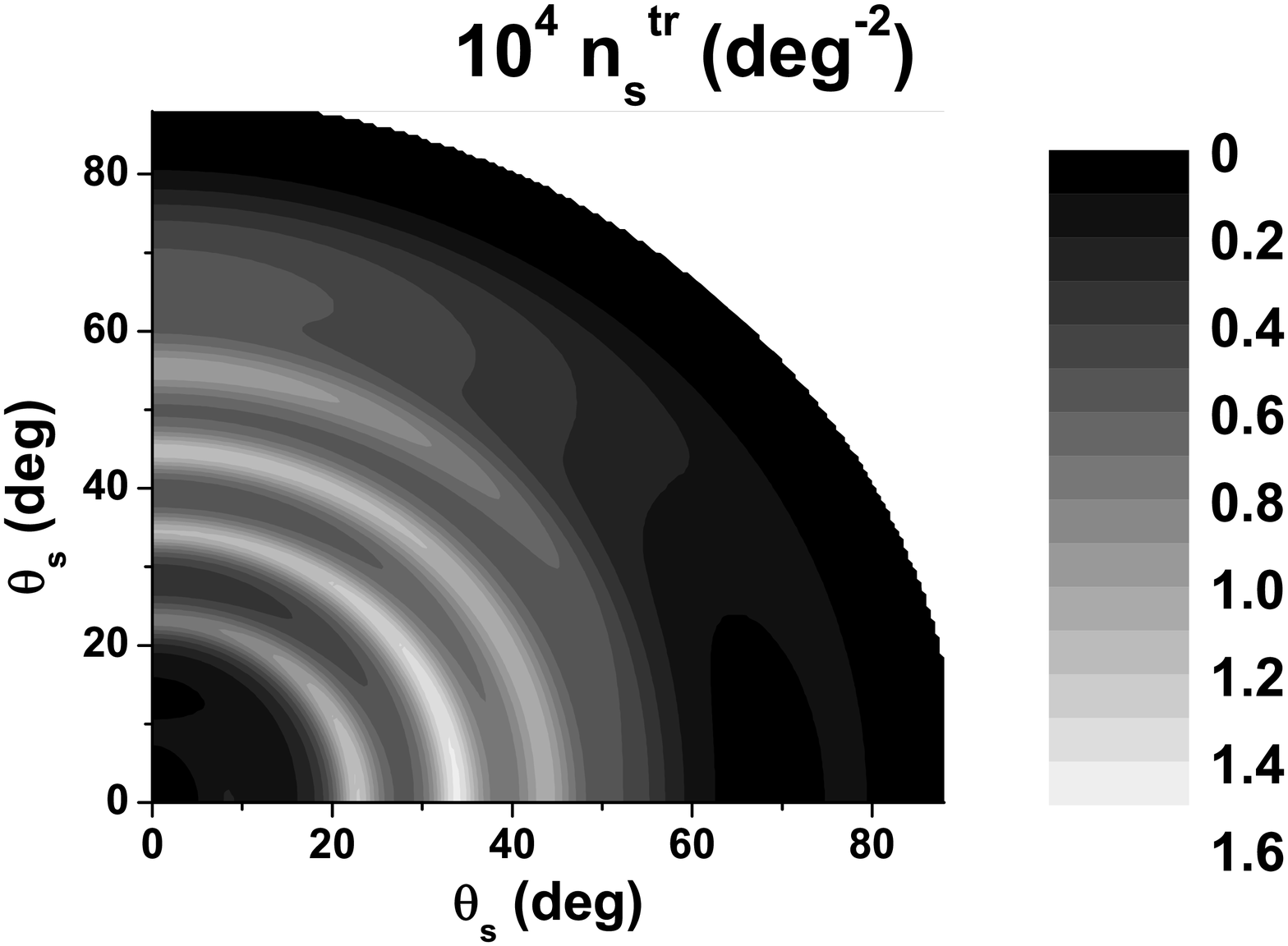}}

 \caption{Densities $ n_s^{\rm tr} $ of mean signal-photon numbers depending
 on signal-field radial ($ \vartheta_s $) and azimuthal ($ \psi_s $)
 emission angles for (a) $ N=11 $, (b) $ N=51 $ and (c) $ N=101 $ layers without
 polarization resolution for both photons propagating forward. The
 emitted signal field is projected onto a hemisphere, one quadrant
 of which is plotted. In the graphs, the radial emission angle $ \vartheta_s $
 determines the distance from the origin, whereas the azimuthal
 emission angle $ \psi_s $ gives the rotation measured from the vertical
 direction. Behavior of the density $ n_s^{\rm tr} $ in the
 remaining three quadrants can be derived from symmetry. The density
 $ n_s^{\rm tr} $
 is normalized such that $ \int_{0}^{\pi/2} \sin(\vartheta_s) d\vartheta_s
 \int_{-\pi/2}^{0} d \psi_s n_s^{\rm tr}(\vartheta_s,\psi_s) =
 (\pi/180)^2/4 $; $ r_p \rightarrow \infty $.}
\label{fig3}
\end{figure}

We consider three examples of layered structures made of GaN/AlN
that demonstrate typical features of the emitted photon pairs.
They generate photon pairs with different polarizations of the
signal and idler photons. They are designed in such a way that
three different mechanisms of splitting the correlated area are
observed. The pump field propagating at normal incidence at the
carrying wavelength $ \lambda_p^0 = 400 $~nm is assumed. The
signal and idler fields are emitted around the degenerate
wavelengths $ \lambda_s^0 = \lambda_i^0 = 800 $~nm in
non-collinear geometry. The GaN and AlN layers are positioned such
that their optical axes are perpendicular to the boundaries. The
considered structures differ in the number $ N $ of layers ($ N=11
$, 51 and 101). This results in different signal-field intensity
transverse profiles plotted in Fig.~\ref{fig3}. Whereas the
shortest structure with $ N=11 $ layers has only one emission
area, the longest one with $ N=101 $ layers already forms five
emission rings. We note that transverse profiles of the
down-converted fields can also be efficiently tailored by
modifying the pump-beam profile \cite{Walborn2004}.

The analyzed structures have certain properties in common. Namely,
they cannot generate proton pairs in nearly collinear geometries
because of the symmetry that cancels the overlap integral. As for
correlated areas, their azimuthal spreads $ \Delta\psi_i $ depend
on the width $ r_p $ of transverse pump-field profile
\cite{Molina-Terriza2005,Hamar2010}. The radial spreads $ \Delta
\vartheta_i $ depend also on the geometry of the layered
structure; the greater the number $ N $ of layers, the smaller the
correlated area. In more detail, the considered structures behave
as follows.

\subsection{Structure with 11 layers}

\begin{figure}    
 \resizebox{0.8\hsize}{!}{\includegraphics{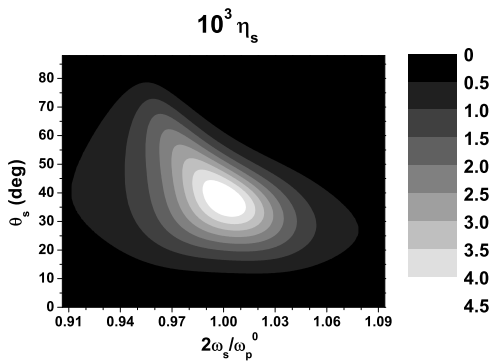}}

 \caption{Relative density $ \eta_s $ of mean signal-photon numbers depending
 on normalized signal-field frequency $ 2\omega_s/\omega_p^0 $ and
 radial emission angle $ \vartheta_s $ for the structure with $ N=11 $
 layers for both photons with arbitrary polarizations propagating forward;
 $ \psi_s = 0 $~deg, $ r_p \rightarrow \infty $.}
\label{fig4}
\end{figure}

The structure is composed of 6 nonlinear GaN layers 90.14~nm long
and 5 linear AlN layers 74.92~nm long. Despite a relatively small
number of layers, photonic band gaps are already formed though the
bottoms of forbidden bands have intensity transmissions around
0.3. For $ \psi_s = 0 $~deg, an efficient SPDC occurs for
polarizations (TE,TM,TE) and (TE,TE,TM) for the
(pump,signal,idler) fields due to the nonlinear coefficient $
d(1,1,3) $. The fields are tuned to the first lower transmission
peaks of the first and second forbidden bands, so the efficiency
of SPDC is nearly optimal. The signal and idler transmission peaks
are spectrally broad and, as a consequence, we observe one large
emission area in the signal-field transverse plane that extends
from cca $ \vartheta_s = 20 $~deg to 60~deg [see Fig.~\ref{3}(a)].
The relative density $ \eta_s $ of mean signal-photon numbers as
plotted in Fig.~\ref{fig4} reveals that the emitted photons have
nearly degenerate frequencies and relatively broad spectra. The
shape of relative density $ \eta_s $ as shown in Fig.~\ref{fig4}
is caused by a different radial dependence of intensity
transmissions $ T_{\rm TE} $ and $ T_{\rm TM} $. In the area
around $ \psi_s = \pm 90 $~deg there occurs no photon-pair
generation because of geometric reasons (the propagation of signal
and idler fields as TM waves is not supported). Compared to one
GaN layer $ 6\times 90.14 $~nm long, the structure gives cca 2
times greater relative densities $ \eta_s $.

\begin{figure}    
 (a) \resizebox{0.8\hsize}{!}{\includegraphics{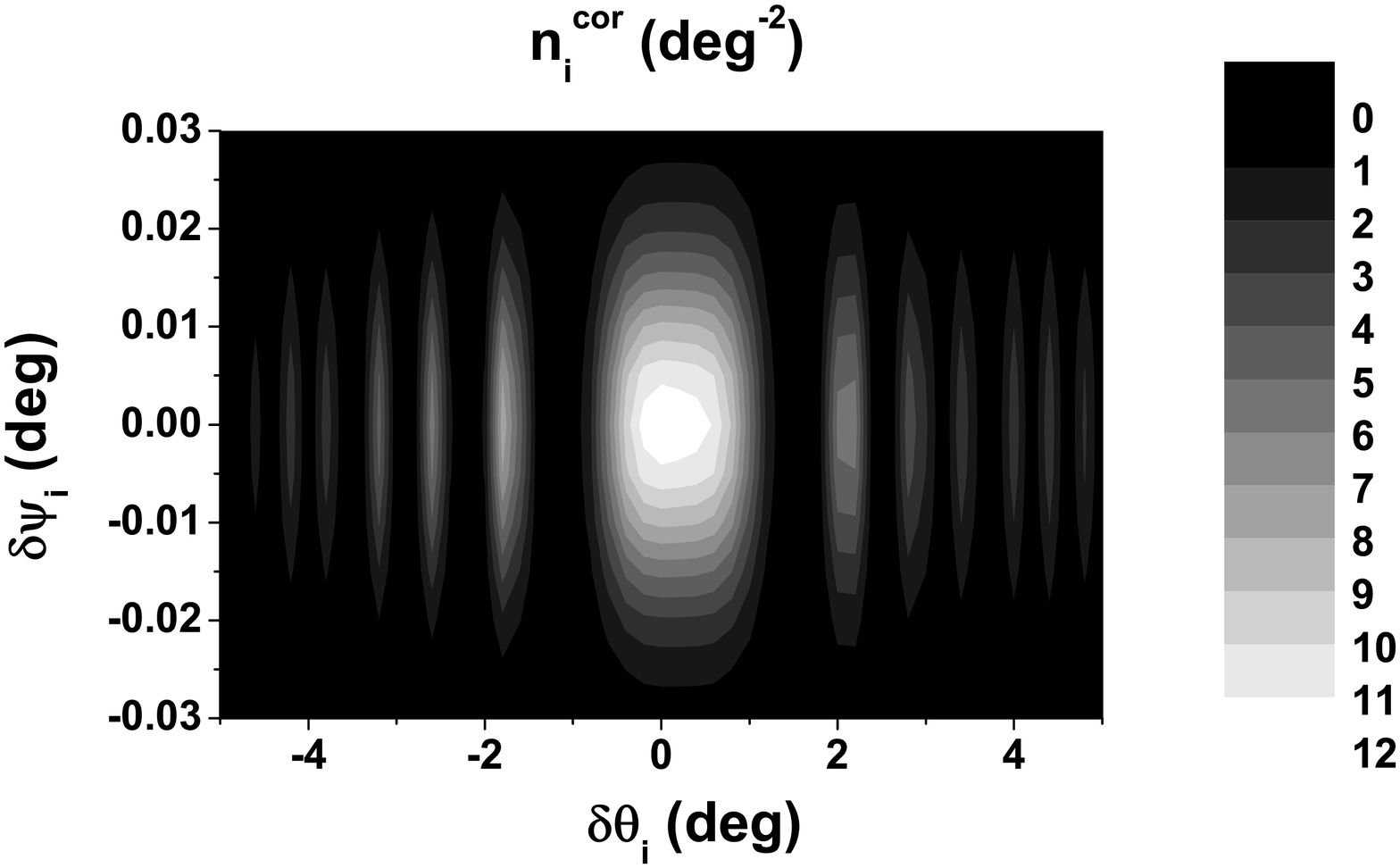}}

 \vspace{2mm}

 (b) \resizebox{0.8\hsize}{!}{\includegraphics{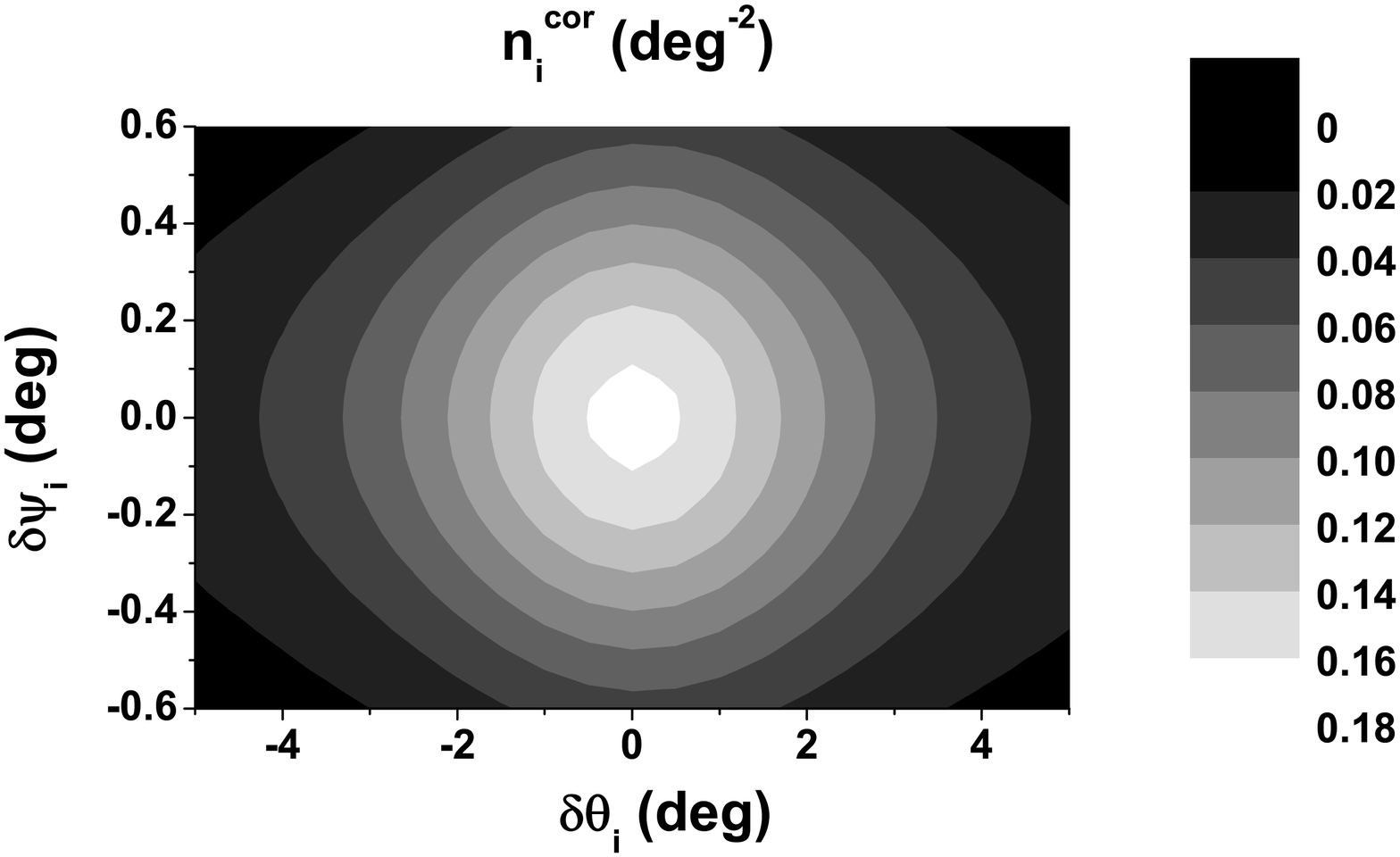}}

 \caption{Correlated area $ n_i^{\rm cor}(\vartheta_i,\psi_i) $
  of an idler photon for (a) $ r_p = 1 $~mm and (b) $ r_p =
  30 $~nm belonging to a signal photon propagating along direction
  $ \vartheta_s^0 = 38 $~deg and $ \psi_s^0 = 0 $~deg. Both photons with arbitrary polarizations
  propagate forward; $ \vartheta_i = \vartheta_i^0 + \delta \vartheta_i $, $ \vartheta_i^0 =
  -\vartheta_s^0 $, $ \psi_i = \psi_i^0 + \delta\psi_i $, $ \psi_i^0 = - \psi_s^0 $.
  The normalization
  $ \int_{-\pi/2}^{\pi/2} d\vartheta_i \int_{-\pi/2}^{\pi/2}
  d\psi_i n_i^{\rm cor}(\vartheta_i,\psi_i)= (\pi/180)^2 $ is used.}
\label{fig5}
\end{figure}

The correlated area of an idler photon for a collimated pump beam
is composed of several 'islands' localized around different
idler-field radial emission angles $ \vartheta_i $ [see
Fig.~\ref{fig5}(a)]. This breaking is caused by the zig-zag
movement of two photons inside the structure after being generated
in one layer. The comparison of correlated areas for collimated
and focused pump beams using, e.g., the graphs in
Fig.~\ref{fig5}(a) and (b) reveals that the spread of correlated
area along the azimuthal emission angle $ \psi_i $ depends
strongly on the amount of focusing the pump beam. On the other
hand, the shape of correlated area in the radial emission angle $
\vartheta_i $ is blurred by a focused pump beam keeping the
overall spread of correlated area roughly the same.

\subsection{Structure with 51 layers}

\begin{figure}    
 \resizebox{0.78\hsize}{!}{\includegraphics{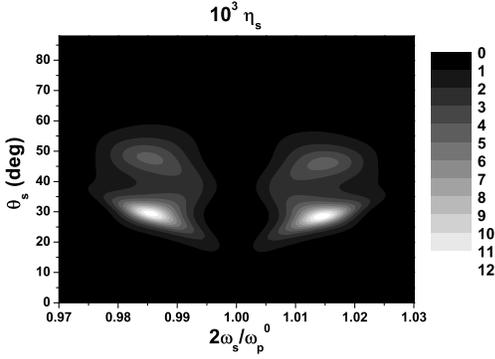}}

 \caption{Relative density $ \eta_s $ of mean signal-photon numbers
 as it depends on normalized signal-field frequency $ 2\omega_s/\omega_p^0 $ and
 radial emission angle $ \vartheta_s $ analyzing the structure with $ N=51 $
 layers and both forward-propagating photons with arbitrary polarizations;
 $ \psi_s = 0 $~deg, $ r_p \rightarrow \infty $.}
\label{fig6}
\end{figure}

The second structure contains 26 nonlinear GaN layers 106.87~nm
long and 25 linear AlN layers 65.99~nm long. This structure has
already well-formed forbidden bands. As Fig.~\ref{fig3}(b) shows
there occurs an efficient SPDC in two concentric rings in the
transverse plane. The structure is designed such that an efficient
nonlinear interaction for fields' polarizations (TM,TM,TM) occurs
along the direction $ \psi_s = 0 $~deg using the nonlinear
coefficient $ d(2,2,3) $. The pump field lies in the first upper
transmission peak of the second forbidden band and the signal and
idler fields are in the first and second lower transmission peaks
of the first forbidden band. The relative density $ \eta_s $ of
mean signal-photon numbers as plotted in Fig.~\ref{fig6}
demonstrates that the signal-field (and also idler-field)
intensity spectra are composed of two symmetric peaks. There
occurs no photon-pair generation for degenerate signal- and
idler-field frequencies because of symmetry. This indicates the
generation of photon pairs in the state that is anti-symmetric
with respect to the exchange of the signal- and idler-field
frequencies. Anti-bunching of photons in a pair and
anti-coalescence (observed in a Hong-Ou-Mandel interferometer) are
the distinguished features of this state \cite{PerinaJr2007b}. The
graph in Fig.~\ref{fig3}(b) reveals that an efficient SPDC occurs
also for azimuthal emission angles $ \psi_s $ around $ \pm 90
$~deg. In this case, the signal photon is generated as TE wave and
the idler photon as TM wave or vice versa. The structure gives cca
50 times greater relative densities $ \eta_s $ in comparison with
one GaN layer $ 26\times 106.87 $~nm long.

\begin{figure}    
 (a) \resizebox{0.8\hsize}{!}{\includegraphics{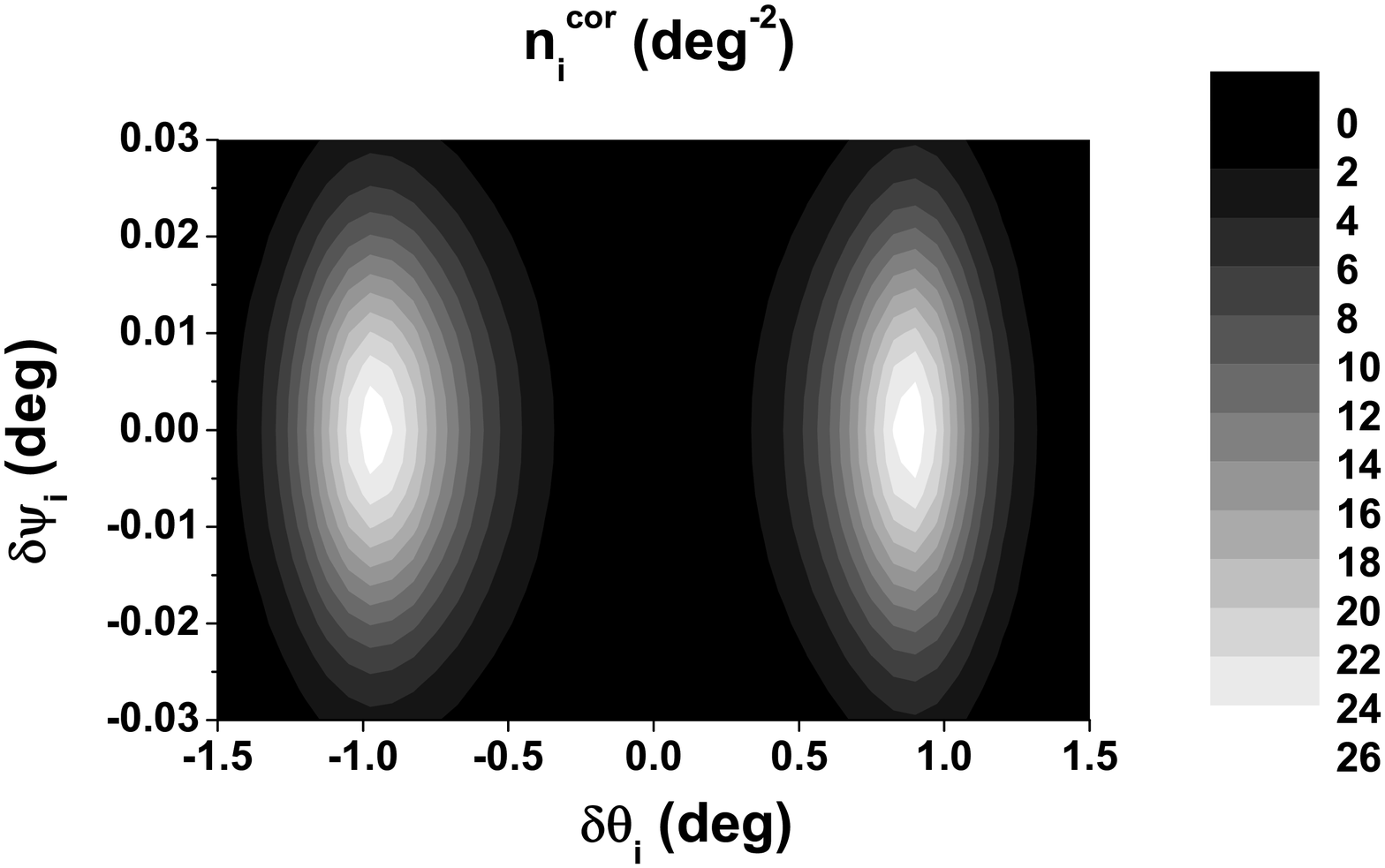}}

 \vspace{2mm}

 (b) \resizebox{0.8\hsize}{!}{\includegraphics{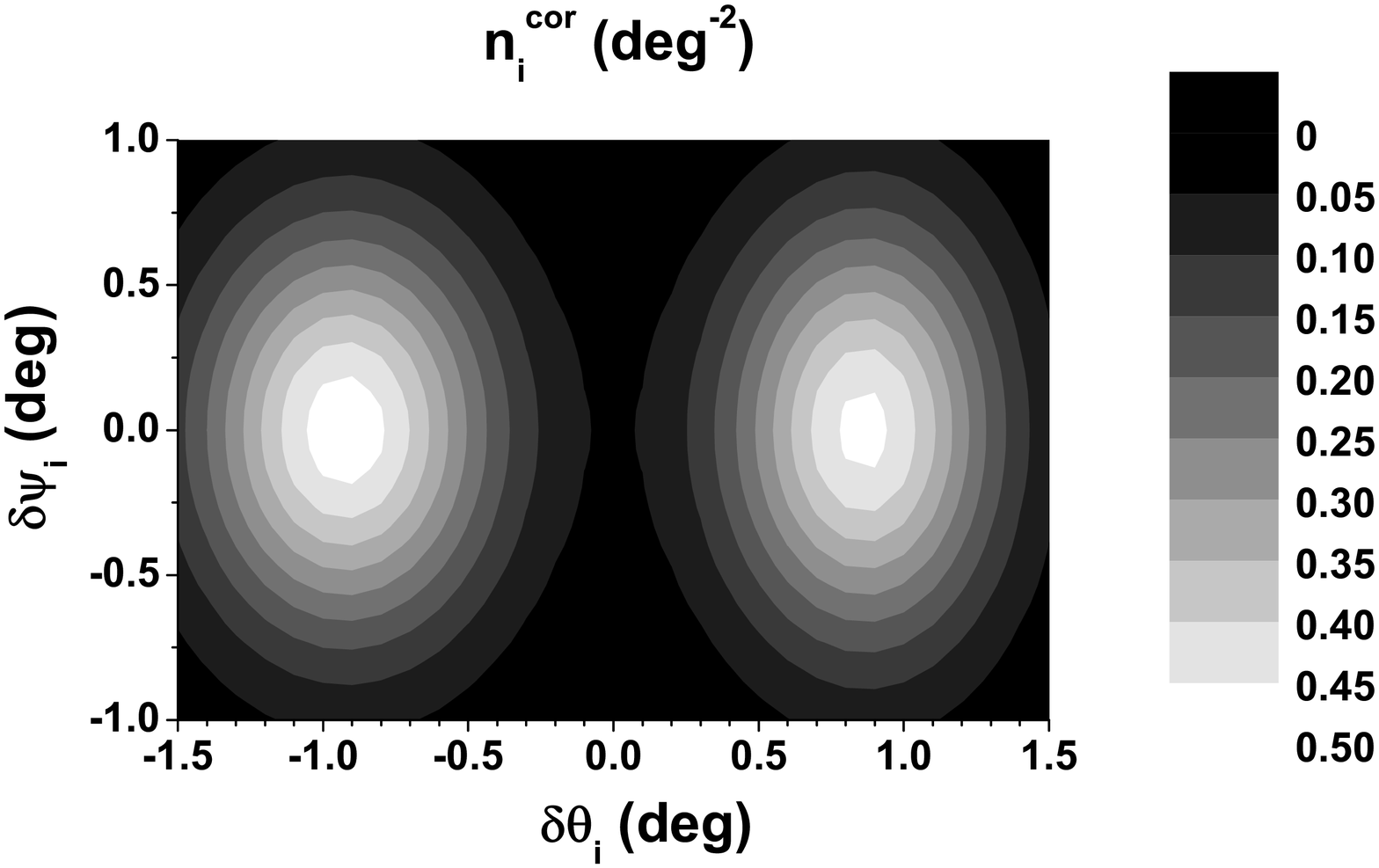}}

 \caption{Correlated area $ n_i^{\rm cor}(\vartheta_i,\psi_i) $ of an
  idler photon for (a) $ r_p = 1 $~mm and (b) $ r_p =
  30 $~nm corresponding to a signal photon propagating along direction
  $ \vartheta_s^0 = 29 $~deg and $ \psi_s^0 = 0 $~deg; $ N=51 $.
  Both photons of arbitrary polarizations
  propagate forward; $ \vartheta_i = \vartheta_i^0 + \delta \vartheta_i $, $ \vartheta_i^0 =
  - \vartheta_s^0 $, $ \psi_i = \psi_i^0 + \delta\psi_i $, $ \psi_i^0 = - \psi_s^0 $.}
\label{fig7}
\end{figure}

As a consequence of two-peak structure of the signal- and
idler-field spectra there occurs splitting of the correlated area
of an idler photon, as documented in Fig.~\ref{fig7}(a). The
overall correlated area is composed of two symmetric parts.
Because this splitting arises from symmetry, it survives even for
focused pump beams, as demonstrated in Fig.~\ref{fig7}(b).

\subsection{Structure with 101 layers}

\begin{figure}    
 \resizebox{0.8\hsize}{!}{\includegraphics{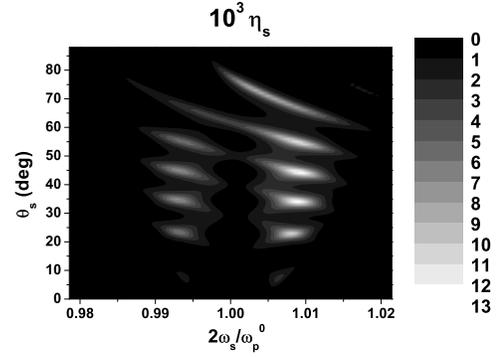}}

 \caption{Relative density $ \eta_s $ of mean signal-photon numbers depending
 on normalized signal-field frequency $ 2\omega_s/\omega_p^0 $ and
 radial emission angle $ \vartheta_s $. The structure with $ N=101 $
 layers and both photons of arbitrary polarizations propagating forward are assumed;
 $ \psi_s = 0 $~deg, $ r_p \rightarrow \infty $.}
\label{fig8}
\end{figure}

The last structure is composed of 51 nonlinear GaN layers
106.42~nm long and 50 linear AlN layers 65.71~nm long. The pump
field lies in the first upper transmission peak of the second
forbidden band. A detailed comparison of the relative density $
\eta_s $ of mean signal-photon numbers (see Fig.~\ref{fig8}) with
intensity transmission spectra $ T_{\rm TE} $ and $ T_{\rm TM} $
valid for TE and TM waves, respectively, and obtained for $ \psi =
0 $~deg (see Fig.~\ref{9}) reveals the following. An efficient
SPDC occurs at intersections of $ j $th and $ (j+1) $th lower
transmissions peaks of the first forbidden band for $ j=2,3,4,5,6
$. This results in five concentric rings in the density $ n_s^{\rm
tr} $ of mean signal-photon numbers clearly visible in
Fig.~\ref{fig3}(c). We note that the nonlinear coefficient $
d(1,1,3) $ is exploited here and (TE,TM,TE) and (TE,TE,TM)
[(TM,TM,TM)] polarizations are suitable for the angles around $
\psi_s = 0 $~deg [$ \psi_s = \pm 90$~deg]. The relative density $
\eta_s $ of mean signal-photon numbers indicates that the
signal-field intensity spectra are composed of two peaks of
different weights for $ \psi_s = 0 $~deg. One peak is related to a
TE signal-field wave, whereas the second one arises from a TM
signal-field wave. Comparison with one GaN layer $ 51\times 106.42
$~nm long reveals that the enhancement of optical fields inside
the structure results in an increase of the relative density $
\eta_s $ cca 330 times.
\begin{figure}    
 (a) \resizebox{0.8\hsize}{!}{\includegraphics{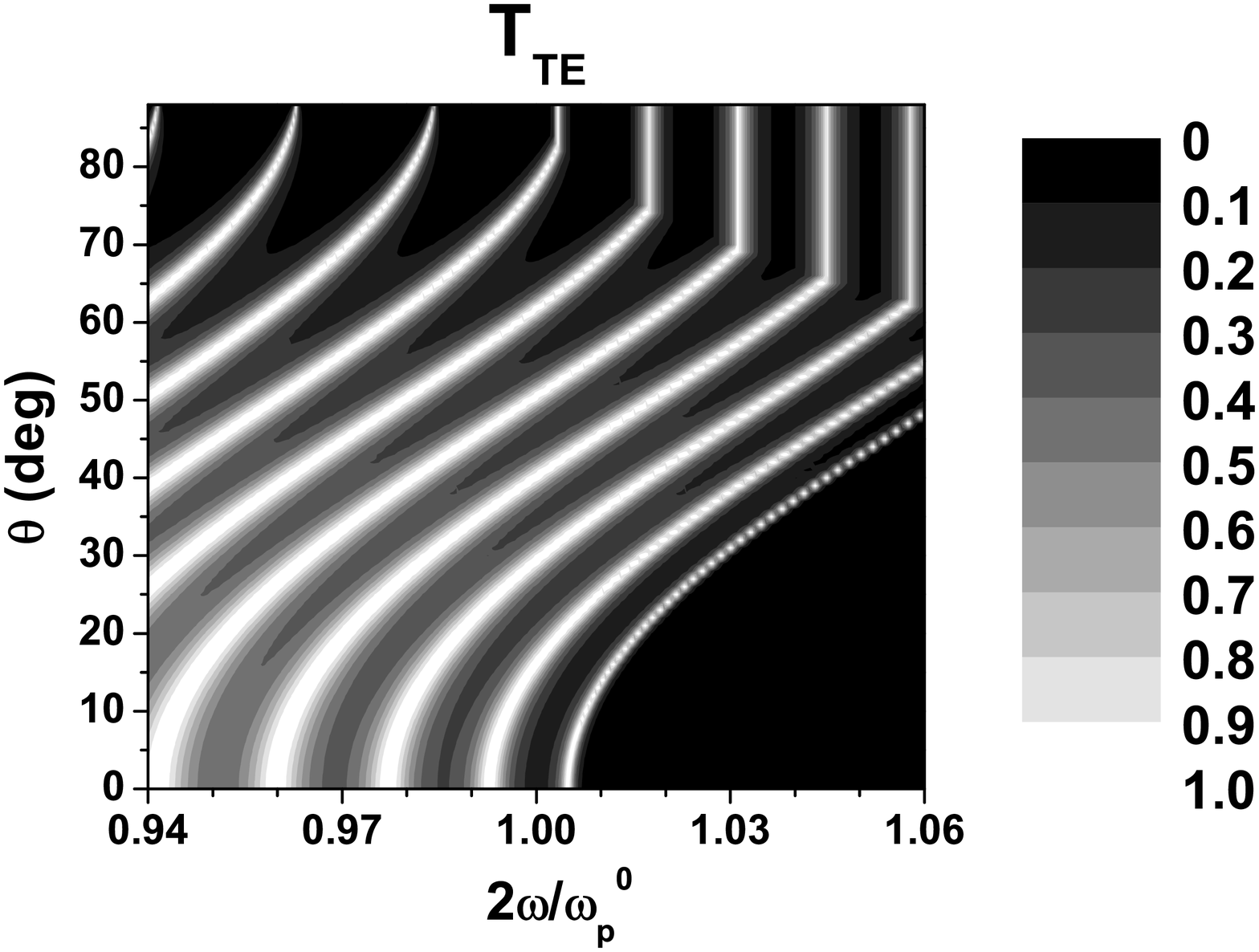}}

 \vspace{2mm}

 (b) \resizebox{0.8\hsize}{!}{\includegraphics{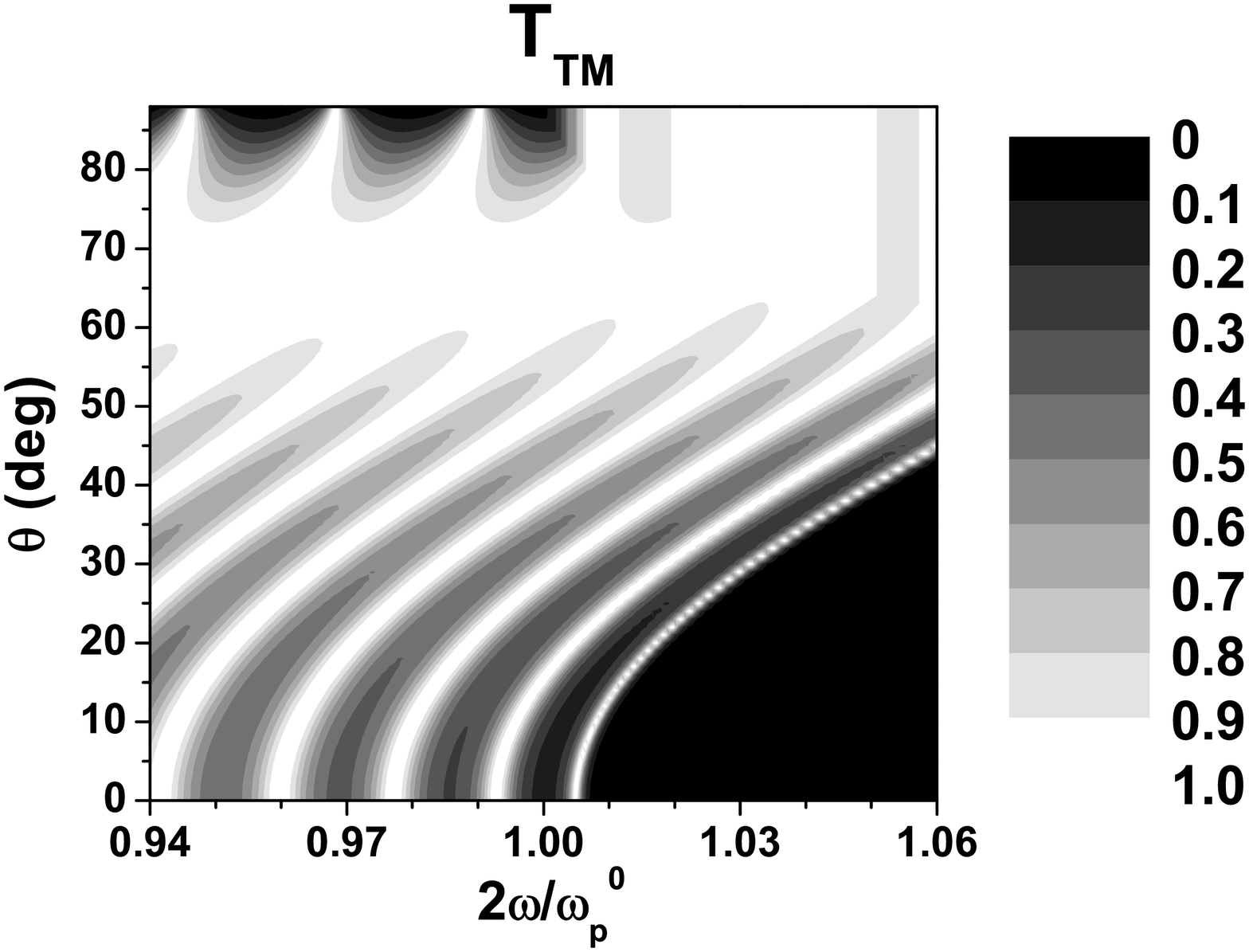}}

 \caption{Intensity transmission coefficients $ T_{\rm TE} $ for TE waves [(a)] and
 $ T_{\rm TM} $ for TM waves [(b)] as they depend
 on radial emission angle $ \vartheta $ assuming $ \psi = 0 $~deg and the structure with $ N=101 $ layers.
 For $ \psi = \pm 90 $~deg, the graphs for TE and TM polarizations are mutually exchanged.}
\label{fig9}
\end{figure}

\begin{figure}    
 (a) \resizebox{0.8\hsize}{!}{\includegraphics{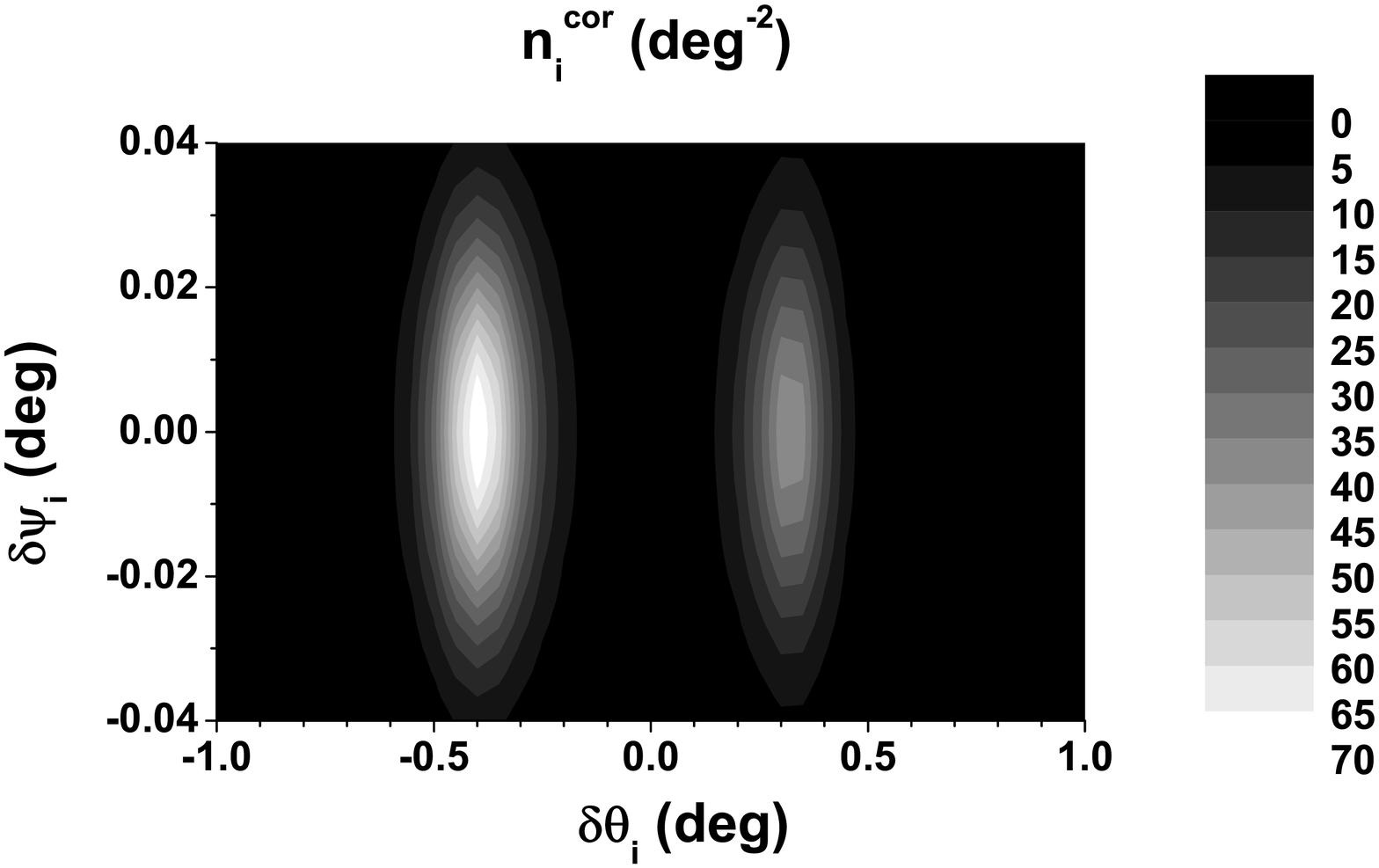}}

 \vspace{2mm}

 (b) \resizebox{0.8\hsize}{!}{\includegraphics{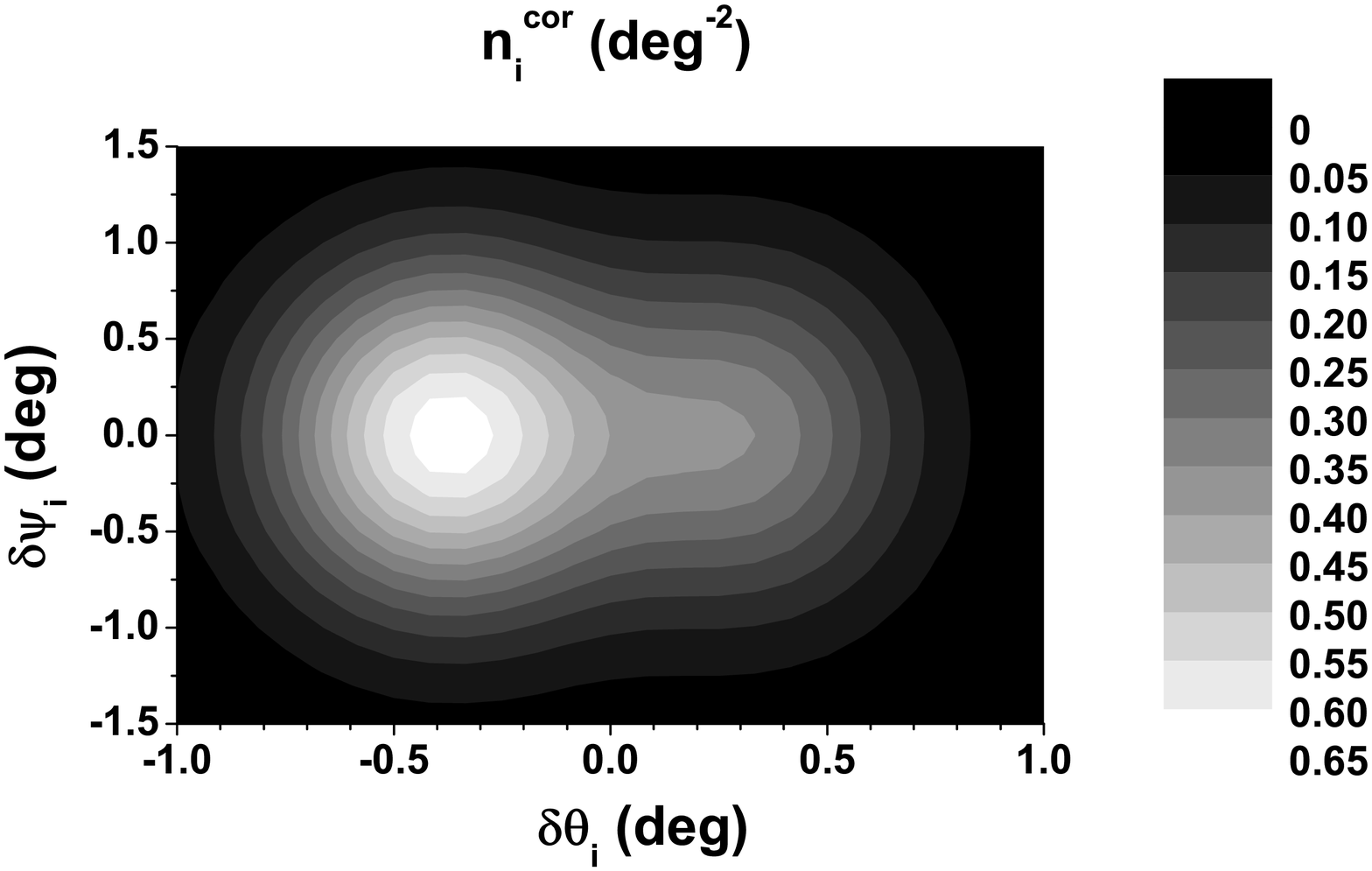}}

 \caption{Correlated area $ n_i^{\rm cor}(\vartheta_i,\psi_i) $ of an idler photon
  for (a) $ r_p = 1 $~mm and (b) $ r_p =
  30 $~nm belonging to a signal photon propagating along direction
  $ \vartheta_s^0 = 23 $~deg and $ \psi_s^0 = 0 $~deg. Both photons have arbitrary polarizations
  and propagate forward; $ \vartheta_i = \vartheta_i^0 + \delta \vartheta_i $, $ \vartheta_i^0 =
  -\vartheta_s^0 $, $ \psi_i = \psi_i^0 + \delta\psi_i $, $ \psi_i^0 = - \psi_s^0 $.}
\label{fig10}
\end{figure}

A two-peak spectral structure leads to splitting of the correlated
area of the idler photon into two parts \cite{PerinaJr2007}, as
shown in Fig.~\ref{fig10}(a). However, two parts are not symmetric
in this case and they can also merge together provided that the
pump beam is sufficiently focused [see Fig.~\ref{fig10}(b)]. The
distance between two parts in the idler-photon radial emission
angle $ \vartheta_i $ increases with the increasing signal-photon
radial emission angle $ \vartheta_s^0 $. Whereas this distance
equals cca 0.6~deg for $ \vartheta_s^0 = 23 $~deg (the first
ring), it equals already cca 4~deg for $ \vartheta_s^0 = 66 $~deg
(the fifth ring). Idler photons found in different parts of the
correlated area differ in their polarizations.

The above investigated structures demonstrate main features of
photon pairs generated in nonlinear layered structures. These
structures allow to generate photon pairs entangled in
frequencies, polarizations as well as in emission directions.
Moreover, they allow tailoring of properties of photon pairs
varying namely the number of layers. A small amount of nonlinear
material inside them is compensated by an increase of
electric-field amplitudes originating in fields' back-scattering.

\section{Conclusions}

We have developed a spatial vectorial quantum model of spontaneous
parametric down-conversion in nonlinear layered structures. Photon
pairs generated in these structures can be entangled in
frequencies, polarizations as well as emission directions. Namely
entanglement in emission directions is important because it can be
efficiently tailored varying the number of layers. A method for
designing efficient layered structures has been suggested. Its
efficiency has been demonstrated suggesting three typical
structures. It has been shown that the number of generated
photon-pairs increases greater than the second power of the number
of layers. Signal-field intensity profiles and correlated areas in
the transverse plane have been analyzed in the designed
structures. Intensity profiles are typically composed of
concentric rings. The greater the number of layers, the greater
the number of rings. Correlated areas may be broken into several
parts because of three possible reasons: i) Zig-zag movement of
photons inside the structure, ii) Necessity to obey geometric
symmetries, and iii) Polarization-dependent properties. Splitting
of the correlated area arising from the geometric symmetry
survives even for a focused pump beam. Also states of photon pairs
showing anti-bunching and anti-coalescence can efficiently be
generated in layered structures. We believe that nonlinear layered
structures are potentially interesting namely as efficient sources
of photon-pair fields entangled in propagation directions which
might be useful, e.g., in 'parallel processing' of quantum
information.

\acknowledgments Support by projects IAA100100713 of GA AV \v{C}R
and COST OC 09026, 1M06002 and Operational Program Research and
Development for Innovations - European Social Fund (project
CZ.1.05/2.1.00/03.0058) of the Ministry of Education of the Czech
Republic is acknowledged. The author thanks M.~Centini for useful
discussions.

\bibliography{perina}

\begin{thebibliography}{52}
\expandafter\ifx\csname natexlab\endcsname\relax\def\natexlab#1{#1}\fi
\expandafter\ifx\csname bibnamefont\endcsname\relax
  \def\bibnamefont#1{#1}\fi
\expandafter\ifx\csname bibfnamefont\endcsname\relax
  \def\bibfnamefont#1{#1}\fi
\expandafter\ifx\csname citenamefont\endcsname\relax
  \def\citenamefont#1{#1}\fi
\expandafter\ifx\csname url\endcsname\relax
  \def\url#1{\texttt{#1}}\fi
\expandafter\ifx\csname urlprefix\endcsname\relax\def\urlprefix{URL }\fi
\providecommand{\bibinfo}[2]{#2}
\providecommand{\eprint}[2][]{\url{#2}}

\bibitem[{\citenamefont{Hong et~al.}(1987)\citenamefont{Hong, Ou, and
  Mandel}}]{Hong1987}
\bibinfo{author}{\bibfnamefont{C.~K.} \bibnamefont{Hong}},
  \bibinfo{author}{\bibfnamefont{Z.~Y.} \bibnamefont{Ou}}, \bibnamefont{and}
  \bibinfo{author}{\bibfnamefont{L.}~\bibnamefont{Mandel}},
  \bibinfo{journal}{Phys. Rev. Lett.} \textbf{\bibinfo{volume}{59}},
  \bibinfo{pages}{2044} (\bibinfo{year}{1987}).

\bibitem[{\citenamefont{Mandel and Wolf}(1995)}]{Mandel1995}
\bibinfo{author}{\bibfnamefont{L.}~\bibnamefont{Mandel}} \bibnamefont{and}
  \bibinfo{author}{\bibfnamefont{E.}~\bibnamefont{Wolf}},
  \emph{\bibinfo{title}{Optical Coherence and Quantum Optics}}
  (\bibinfo{publisher}{Cambridge Univ. Press, Cambridge},
  \bibinfo{year}{1995}).

\bibitem[{Bou(2000)}]{Bouwmeester2000}
in \emph{\bibinfo{booktitle}{The Physics of Quantum Information}}, edited by
  \bibinfo{editor}{\bibfnamefont{D.}~\bibnamefont{Bouwmeester}},
  \bibinfo{editor}{\bibfnamefont{A.}~\bibnamefont{Ekert}}, \bibnamefont{and}
  \bibinfo{editor}{\bibfnamefont{A.}~\bibnamefont{Zeilinger}}
  (\bibinfo{publisher}{Springer, Berlin}, \bibinfo{year}{2000}).

\bibitem[{\citenamefont{Pe\v{r}ina et~al.}(1994)\citenamefont{Pe\v{r}ina,
  Hradil, and Jur\v{c}o}}]{Perina1994}
\bibinfo{author}{\bibfnamefont{J.}~\bibnamefont{Pe\v{r}ina}},
  \bibinfo{author}{\bibfnamefont{Z.}~\bibnamefont{Hradil}}, \bibnamefont{and}
  \bibinfo{author}{\bibfnamefont{B.}~\bibnamefont{Jur\v{c}o}},
  \emph{\bibinfo{title}{Quantum Optics and Fundamentals of Physics}}
  (\bibinfo{publisher}{Kluwer, Dordrecht}, \bibinfo{year}{1994}).

\bibitem[{\citenamefont{Bouwmeester et~al.}(1997)\citenamefont{Bouwmeester,
  Pan, Mattle, Eibl, Weinfurter, and Zeilinger}}]{Bouwmeester1997}
\bibinfo{author}{\bibfnamefont{D.}~\bibnamefont{Bouwmeester}},
  \bibinfo{author}{\bibfnamefont{J.~W.} \bibnamefont{Pan}},
  \bibinfo{author}{\bibfnamefont{K.}~\bibnamefont{Mattle}},
  \bibinfo{author}{\bibfnamefont{M.}~\bibnamefont{Eibl}},
  \bibinfo{author}{\bibfnamefont{H.}~\bibnamefont{Weinfurter}},
  \bibnamefont{and}
  \bibinfo{author}{\bibfnamefont{A.}~\bibnamefont{Zeilinger}},
  \bibinfo{journal}{Nature} \textbf{\bibinfo{volume}{390}},
  \bibinfo{pages}{575} (\bibinfo{year}{1997}).

\bibitem[{\citenamefont{Harris}(2007)}]{Harris2007}
\bibinfo{author}{\bibfnamefont{S.~E.} \bibnamefont{Harris}},
  \bibinfo{journal}{Phys. Rev. Lett.} \textbf{\bibinfo{volume}{98}},
  \bibinfo{eid}{063602} (\bibinfo{year}{2007}).

\bibitem[{\citenamefont{Brida et~al.}(2009)\citenamefont{Brida, Chekhova,
  Degiovanni, Genovese, Kitaeva, Meda, and Shumilkina}}]{Brida2009}
\bibinfo{author}{\bibfnamefont{G.}~\bibnamefont{Brida}},
  \bibinfo{author}{\bibfnamefont{M.~V.} \bibnamefont{Chekhova}},
  \bibinfo{author}{\bibfnamefont{I.~P.} \bibnamefont{Degiovanni}},
  \bibinfo{author}{\bibfnamefont{M.}~\bibnamefont{Genovese}},
  \bibinfo{author}{\bibfnamefont{G.~K.} \bibnamefont{Kitaeva}},
  \bibinfo{author}{\bibfnamefont{A.}~\bibnamefont{Meda}}, \bibnamefont{and}
  \bibinfo{author}{\bibfnamefont{O.~A.} \bibnamefont{Shumilkina}},
  \bibinfo{journal}{Phys. Rev. Lett.} \textbf{\bibinfo{volume}{103}},
  \bibinfo{pages}{193602} (\bibinfo{year}{2009}).

\bibitem[{\citenamefont{Keller and Rubin}(1997)}]{Keller1997}
\bibinfo{author}{\bibfnamefont{T.~E.} \bibnamefont{Keller}} \bibnamefont{and}
  \bibinfo{author}{\bibfnamefont{M.~H.} \bibnamefont{Rubin}},
  \bibinfo{journal}{Phys. Rev. A} \textbf{\bibinfo{volume}{56}},
  \bibinfo{pages}{1534} (\bibinfo{year}{1997}).

\bibitem[{\citenamefont{{Pe\v{r}ina~Jr.}
  et~al.}(1999)\citenamefont{{Pe\v{r}ina~Jr.}, Sergienko, Jost, Saleh, and
  Teich}}]{PerinaJr1999a}
\bibinfo{author}{\bibfnamefont{J.}~\bibnamefont{{Pe\v{r}ina~Jr.}}},
  \bibinfo{author}{\bibfnamefont{A.~V.} \bibnamefont{Sergienko}},
  \bibinfo{author}{\bibfnamefont{B.~M.} \bibnamefont{Jost}},
  \bibinfo{author}{\bibfnamefont{B.~E.~A.} \bibnamefont{Saleh}},
  \bibnamefont{and} \bibinfo{author}{\bibfnamefont{M.~C.} \bibnamefont{Teich}},
  \bibinfo{journal}{Phys. Rev. A} \textbf{\bibinfo{volume}{59}},
  \bibinfo{pages}{2359} (\bibinfo{year}{1999}).

\bibitem[{\citenamefont{Joobeur et~al.}(1994)\citenamefont{Joobeur, Saleh, and
  Teich}}]{Jobeur1994}
\bibinfo{author}{\bibfnamefont{A.}~\bibnamefont{Joobeur}},
  \bibinfo{author}{\bibfnamefont{B.~E.~A.} \bibnamefont{Saleh}},
  \bibnamefont{and} \bibinfo{author}{\bibfnamefont{M.~C.} \bibnamefont{Teich}},
  \bibinfo{journal}{Physical Review A} \textbf{\bibinfo{volume}{50}},
  \bibinfo{pages}{3349} (\bibinfo{year}{1994}).

\bibitem[{\citenamefont{Joobeur et~al.}(1996)\citenamefont{Joobeur, Saleh,
  Larchuk, and Teich}}]{Jobeur1996}
\bibinfo{author}{\bibfnamefont{A.}~\bibnamefont{Joobeur}},
  \bibinfo{author}{\bibfnamefont{B.~E.~A.} \bibnamefont{Saleh}},
  \bibinfo{author}{\bibfnamefont{T.~S.} \bibnamefont{Larchuk}},
  \bibnamefont{and} \bibinfo{author}{\bibfnamefont{M.~C.} \bibnamefont{Teich}},
  \bibinfo{journal}{Physical Review A} \textbf{\bibinfo{volume}{53}},
  \bibinfo{pages}{4360} (\bibinfo{year}{1996}).

\bibitem[{\citenamefont{Vallone et~al.}(2007)\citenamefont{Vallone, Pomarico,
  Mataloni, {De~Martini}, and Berardi}}]{Vallone2007}
\bibinfo{author}{\bibfnamefont{G.}~\bibnamefont{Vallone}},
  \bibinfo{author}{\bibfnamefont{E.}~\bibnamefont{Pomarico}},
  \bibinfo{author}{\bibfnamefont{P.}~\bibnamefont{Mataloni}},
  \bibinfo{author}{\bibfnamefont{F.}~\bibnamefont{{De~Martini}}},
  \bibnamefont{and} \bibinfo{author}{\bibfnamefont{V.}~\bibnamefont{Berardi}},
  \bibinfo{journal}{Phys. Rev. Lett.} \textbf{\bibinfo{volume}{98}},
  \bibinfo{pages}{180502} (\bibinfo{year}{2007}).

\bibitem[{\citenamefont{Monken et~al.}(1998)\citenamefont{Monken,
  {Souto~Ribeiro}, and Padua}}]{Monken1998}
\bibinfo{author}{\bibfnamefont{C.~H.} \bibnamefont{Monken}},
  \bibinfo{author}{\bibfnamefont{P.~H.} \bibnamefont{{Souto~Ribeiro}}},
  \bibnamefont{and} \bibinfo{author}{\bibfnamefont{S.}~\bibnamefont{Padua}},
  \bibinfo{journal}{Phys. Rev. A} \textbf{\bibinfo{volume}{57}},
  \bibinfo{pages}{3123} (\bibinfo{year}{1998}).

\bibitem[{\citenamefont{Walborn et~al.}(2004)\citenamefont{Walborn,
  {de~Oliveira}, Thebaldi, and Monken}}]{Walborn2004}
\bibinfo{author}{\bibfnamefont{S.~P.} \bibnamefont{Walborn}},
  \bibinfo{author}{\bibfnamefont{A.~N.} \bibnamefont{{de~Oliveira}}},
  \bibinfo{author}{\bibfnamefont{R.~S.} \bibnamefont{Thebaldi}},
  \bibnamefont{and} \bibinfo{author}{\bibfnamefont{C.~H.}
  \bibnamefont{Monken}}, \bibinfo{journal}{Phys. Rev. A}
  \textbf{\bibinfo{volume}{69}}, \bibinfo{pages}{023811}
  (\bibinfo{year}{2004}).

\bibitem[{\citenamefont{Law and Eberly}(2004)}]{Law2004}
\bibinfo{author}{\bibfnamefont{C.~K.} \bibnamefont{Law}} \bibnamefont{and}
  \bibinfo{author}{\bibfnamefont{J.~H.} \bibnamefont{Eberly}},
  \bibinfo{journal}{Phys. Rev. Lett.} \textbf{\bibinfo{volume}{92}},
  \bibinfo{eid}{127903} (\bibinfo{year}{2004}).

\bibitem[{\citenamefont{Mair et~al.}(2001)\citenamefont{Mair, Vaziri, Weihs,
  and Zeilinger}}]{Mair2001}
\bibinfo{author}{\bibfnamefont{A.}~\bibnamefont{Mair}},
  \bibinfo{author}{\bibfnamefont{A.}~\bibnamefont{Vaziri}},
  \bibinfo{author}{\bibfnamefont{G.}~\bibnamefont{Weihs}}, \bibnamefont{and}
  \bibinfo{author}{\bibfnamefont{A.}~\bibnamefont{Zeilinger}},
  \bibinfo{journal}{Nature} \textbf{\bibinfo{volume}{412}},
  \bibinfo{pages}{313} (\bibinfo{year}{2001}).

\bibitem[{\citenamefont{Oemrawsingh et~al.}(2005)\citenamefont{Oemrawsingh, Ma,
  Voigt, Aiello, Eliel, {'t~Hooft}, and Woerdman}}]{Oemrawsingh2005}
\bibinfo{author}{\bibfnamefont{S.~S.~R.} \bibnamefont{Oemrawsingh}},
  \bibinfo{author}{\bibfnamefont{X.}~\bibnamefont{Ma}},
  \bibinfo{author}{\bibfnamefont{D.}~\bibnamefont{Voigt}},
  \bibinfo{author}{\bibfnamefont{A.}~\bibnamefont{Aiello}},
  \bibinfo{author}{\bibfnamefont{E.~R.} \bibnamefont{Eliel}},
  \bibinfo{author}{\bibfnamefont{G.~W.} \bibnamefont{{'t~Hooft}}},
  \bibnamefont{and} \bibinfo{author}{\bibfnamefont{J.~P.}
  \bibnamefont{Woerdman}}, \bibinfo{journal}{Phys. Rev. Lett.}
  \textbf{\bibinfo{volume}{95}}, \bibinfo{pages}{240501}
  (\bibinfo{year}{2005}).

\bibitem[{\citenamefont{Molina-Terriza
  et~al.}(2005)\citenamefont{Molina-Terriza, Minardi, Deyanova, Osorio,
  Hendrych, and Torres}}]{Molina-Terriza2005}
\bibinfo{author}{\bibfnamefont{G.}~\bibnamefont{Molina-Terriza}},
  \bibinfo{author}{\bibfnamefont{S.}~\bibnamefont{Minardi}},
  \bibinfo{author}{\bibfnamefont{Y.}~\bibnamefont{Deyanova}},
  \bibinfo{author}{\bibfnamefont{C.~I.} \bibnamefont{Osorio}},
  \bibinfo{author}{\bibfnamefont{M.}~\bibnamefont{Hendrych}}, \bibnamefont{and}
  \bibinfo{author}{\bibfnamefont{J.~P.} \bibnamefont{Torres}},
  \bibinfo{journal}{Phys. Rev. A} \textbf{\bibinfo{volume}{72}},
  \bibinfo{pages}{065802} (\bibinfo{year}{2005}).

\bibitem[{\citenamefont{Jost et~al.}(1998)\citenamefont{Jost, Sergienko,
  Abouraddy, Saleh, and Teich}}]{Jost1998}
\bibinfo{author}{\bibfnamefont{B.~M.} \bibnamefont{Jost}},
  \bibinfo{author}{\bibfnamefont{A.~V.} \bibnamefont{Sergienko}},
  \bibinfo{author}{\bibfnamefont{A.~F.} \bibnamefont{Abouraddy}},
  \bibinfo{author}{\bibfnamefont{B.~E.~A.} \bibnamefont{Saleh}},
  \bibnamefont{and} \bibinfo{author}{\bibfnamefont{M.~C.} \bibnamefont{Teich}},
  \bibinfo{journal}{Opt. Express} \textbf{\bibinfo{volume}{3}},
  \bibinfo{pages}{81} (\bibinfo{year}{1998}).

\bibitem[{\citenamefont{Haderka et~al.}(2005)\citenamefont{Haderka,
  {Pe\v{r}ina~Jr.}, and Hamar}}]{Haderka2005}
\bibinfo{author}{\bibfnamefont{O.}~\bibnamefont{Haderka}},
  \bibinfo{author}{\bibfnamefont{J.}~\bibnamefont{{Pe\v{r}ina~Jr.}}},
  \bibnamefont{and} \bibinfo{author}{\bibfnamefont{M.}~\bibnamefont{Hamar}},
  \bibinfo{journal}{J. Opt. B: Quantum Semiclass. Opt.}
  \textbf{\bibinfo{volume}{7}}, \bibinfo{pages}{S572} (\bibinfo{year}{2005}).

\bibitem[{\citenamefont{Rubin and Shih}(2008)}]{Rubin2008}
\bibinfo{author}{\bibfnamefont{M.~H.} \bibnamefont{Rubin}} \bibnamefont{and}
  \bibinfo{author}{\bibfnamefont{Y.~H.} \bibnamefont{Shih}},
  \bibinfo{journal}{Phys. Rev. A} \textbf{\bibinfo{volume}{78}},
  \bibinfo{pages}{033836} (\bibinfo{year}{2008}).

\bibitem[{\citenamefont{Brambilla et~al.}(2004)\citenamefont{Brambilla, Gatti,
  Bache, and Lugiato}}]{Brambilla2004}
\bibinfo{author}{\bibfnamefont{E.}~\bibnamefont{Brambilla}},
  \bibinfo{author}{\bibfnamefont{A.}~\bibnamefont{Gatti}},
  \bibinfo{author}{\bibfnamefont{M.}~\bibnamefont{Bache}}, \bibnamefont{and}
  \bibinfo{author}{\bibfnamefont{L.~A.} \bibnamefont{Lugiato}},
  \bibinfo{journal}{Phys. Rev. A} \textbf{\bibinfo{volume}{69}},
  \bibinfo{pages}{023802} (\bibinfo{year}{2004}).

\bibitem[{\citenamefont{Jedrkiewicz et~al.}(2004)\citenamefont{Jedrkiewicz,
  Jiang, Brambilla, Gatti, Bache, Lugiato, and {Di~Trapani}}}]{Jedrkiewicz2004}
\bibinfo{author}{\bibfnamefont{O.}~\bibnamefont{Jedrkiewicz}},
  \bibinfo{author}{\bibfnamefont{Y.~K.} \bibnamefont{Jiang}},
  \bibinfo{author}{\bibfnamefont{E.}~\bibnamefont{Brambilla}},
  \bibinfo{author}{\bibfnamefont{A.}~\bibnamefont{Gatti}},
  \bibinfo{author}{\bibfnamefont{M.}~\bibnamefont{Bache}},
  \bibinfo{author}{\bibfnamefont{L.~A.} \bibnamefont{Lugiato}},
  \bibnamefont{and}
  \bibinfo{author}{\bibfnamefont{P.}~\bibnamefont{{Di~Trapani}}},
  \bibinfo{journal}{Phys. Rev. Lett.} \textbf{\bibinfo{volume}{93}},
  \bibinfo{pages}{243601} (\bibinfo{year}{2004}).

\bibitem[{\citenamefont{Migdall}(1999)}]{Migdall1999}
\bibinfo{author}{\bibfnamefont{A.}~\bibnamefont{Migdall}},
  \bibinfo{journal}{Physics Today} \textbf{\bibinfo{volume}{41}},
  \bibinfo{pages}{1} (\bibinfo{year}{1999}).

\bibitem[{\citenamefont{Bru\ss{} and {L\"{u}tkenhaus}}(2000)}]{Lutkenhaus2000}
\bibinfo{author}{\bibfnamefont{D.}~\bibnamefont{Bru\ss{}}} \bibnamefont{and}
  \bibinfo{author}{\bibfnamefont{N.}~\bibnamefont{{L\"{u}tkenhaus}}}, in
  \emph{\bibinfo{booktitle}{Applicable Algebra in Engineering, Communication
  and Computing, Vol. 10}} (\bibinfo{publisher}{Springer, Berlin},
  \bibinfo{year}{2000}), p. \bibinfo{pages}{383}.

\bibitem[{\citenamefont{Kitaeva}(2007)}]{Kitaeva2007}
\bibinfo{author}{\bibfnamefont{G.~K.} \bibnamefont{Kitaeva}},
  \bibinfo{journal}{Phys. Rev. A} \textbf{\bibinfo{volume}{76}},
  \bibinfo{eid}{043841} (\bibinfo{year}{2007}).

\bibitem[{\citenamefont{Svozil{\'i}k and
  {Pe\v{r}ina~Jr.}}(2009)}]{Svozilik2009}
\bibinfo{author}{\bibfnamefont{J.}~\bibnamefont{Svozil{\'i}k}}
  \bibnamefont{and}
  \bibinfo{author}{\bibfnamefont{J.}~\bibnamefont{{Pe\v{r}ina~Jr.}}},
  \bibinfo{journal}{Phys. Rev. A} \textbf{\bibinfo{volume}{80}},
  \bibinfo{eid}{023819} (\bibinfo{year}{2009}).

\bibitem[{\citenamefont{Svozil{\'i}k and
  {Pe\v{r}ina~Jr.}}(2010)}]{Svozilik2010}
\bibinfo{author}{\bibfnamefont{J.}~\bibnamefont{Svozil{\'i}k}}
  \bibnamefont{and}
  \bibinfo{author}{\bibfnamefont{J.}~\bibnamefont{{Pe\v{r}ina~Jr.}}},
  \bibinfo{journal}{Opt. Express} \textbf{\bibinfo{volume}{18}},
  \bibinfo{pages}{27130} (\bibinfo{year}{2010}).

\bibitem[{\citenamefont{{U'Ren} et~al.}(2004)\citenamefont{{U'Ren}, Silberhorn,
  Banaszek, and Walmsley}}]{URen2004}
\bibinfo{author}{\bibfnamefont{A.~B.} \bibnamefont{{U'Ren}}},
  \bibinfo{author}{\bibfnamefont{C.}~\bibnamefont{Silberhorn}},
  \bibinfo{author}{\bibfnamefont{K.}~\bibnamefont{Banaszek}}, \bibnamefont{and}
  \bibinfo{author}{\bibfnamefont{I.~A.} \bibnamefont{Walmsley}},
  \bibinfo{journal}{Phys. Rev. Lett.} \textbf{\bibinfo{volume}{93}},
  \bibinfo{pages}{093601} (\bibinfo{year}{2004}).

\bibitem[{\citenamefont{Spillane et~al.}(2007)\citenamefont{Spillane,
  Fiorentino, and Beausoleil}}]{Spillane2007}
\bibinfo{author}{\bibfnamefont{S.~M.} \bibnamefont{Spillane}},
  \bibinfo{author}{\bibfnamefont{M.}~\bibnamefont{Fiorentino}},
  \bibnamefont{and} \bibinfo{author}{\bibfnamefont{R.~G.}
  \bibnamefont{Beausoleil}}, \bibinfo{journal}{Opt. Express}
  \textbf{\bibinfo{volume}{15}}, \bibinfo{pages}{8770} (\bibinfo{year}{2007}).

\bibitem[{\citenamefont{Chen et~al.}(2009)\citenamefont{Chen, Pearlman, Ling,
  Fan, and Migdall}}]{Chen2009}
\bibinfo{author}{\bibfnamefont{J.}~\bibnamefont{Chen}},
  \bibinfo{author}{\bibfnamefont{A.~J.} \bibnamefont{Pearlman}},
  \bibinfo{author}{\bibfnamefont{A.}~\bibnamefont{Ling}},
  \bibinfo{author}{\bibfnamefont{J.}~\bibnamefont{Fan}}, \bibnamefont{and}
  \bibinfo{author}{\bibfnamefont{A.}~\bibnamefont{Migdall}},
  \bibinfo{journal}{Opt. Express} \textbf{\bibinfo{volume}{17}},
  \bibinfo{pages}{6727} (\bibinfo{year}{2009}).

\bibitem[{Ber(2001)}]{Bertolotti2001}
in \emph{\bibinfo{booktitle}{Nanoscale Linear and Nonlinear Optics, AIP Vol.
  560}}, edited by
  \bibinfo{editor}{\bibfnamefont{M.}~\bibnamefont{Bertolotti}},
  \bibinfo{editor}{\bibfnamefont{C.~M.} \bibnamefont{Bowden}},
  \bibnamefont{and} \bibinfo{editor}{\bibfnamefont{C.}~\bibnamefont{Sibilia}}
  (\bibinfo{publisher}{AIP, Melville}, \bibinfo{year}{2001}).

\bibitem[{\citenamefont{Yablonovitch}(1987)}]{Yablonovitch1987}
\bibinfo{author}{\bibfnamefont{E.}~\bibnamefont{Yablonovitch}},
  \bibinfo{journal}{Phys. Rev. Lett.} \textbf{\bibinfo{volume}{58}},
  \bibinfo{pages}{2059} (\bibinfo{year}{1987}).

\bibitem[{\citenamefont{John}(1987)}]{John1987}
\bibinfo{author}{\bibfnamefont{S.}~\bibnamefont{John}}, \bibinfo{journal}{Phys.
  Rev. Lett.} \textbf{\bibinfo{volume}{58}}, \bibinfo{pages}{2486}
  (\bibinfo{year}{1987}).

\bibitem[{\citenamefont{Vamivakas et~al.}(2004)\citenamefont{Vamivakas, Saleh,
  Sergienko, and Teich}}]{Vamivakas2004}
\bibinfo{author}{\bibfnamefont{A.~N.} \bibnamefont{Vamivakas}},
  \bibinfo{author}{\bibfnamefont{B.~E.~A.} \bibnamefont{Saleh}},
  \bibinfo{author}{\bibfnamefont{A.~V.} \bibnamefont{Sergienko}},
  \bibnamefont{and} \bibinfo{author}{\bibfnamefont{M.~C.} \bibnamefont{Teich}},
  \bibinfo{journal}{Phys. Rev. A} \textbf{\bibinfo{volume}{70}},
  \bibinfo{pages}{043810} (\bibinfo{year}{2004}).

\bibitem[{\citenamefont{Centini et~al.}(2005)\citenamefont{Centini,
  {Pe\v{r}ina~Jr.}, Sciscione, Sibilia, Scalora, Bloemer, and
  Bertolotti}}]{Centini2005}
\bibinfo{author}{\bibfnamefont{M.}~\bibnamefont{Centini}},
  \bibinfo{author}{\bibfnamefont{J.}~\bibnamefont{{Pe\v{r}ina~Jr.}}},
  \bibinfo{author}{\bibfnamefont{L.}~\bibnamefont{Sciscione}},
  \bibinfo{author}{\bibfnamefont{C.}~\bibnamefont{Sibilia}},
  \bibinfo{author}{\bibfnamefont{M.}~\bibnamefont{Scalora}},
  \bibinfo{author}{\bibfnamefont{M.~J.} \bibnamefont{Bloemer}},
  \bibnamefont{and}
  \bibinfo{author}{\bibfnamefont{M.}~\bibnamefont{Bertolotti}},
  \bibinfo{journal}{Phys. Rev. A} \textbf{\bibinfo{volume}{72}},
  \bibinfo{eid}{033806} (\bibinfo{year}{2005}).

\bibitem[{\citenamefont{{Pe\v{r}ina~Jr.}
  et~al.}(2006)\citenamefont{{Pe\v{r}ina~Jr.}, Centini, Sibilia, Bertolotti,
  and Scalora}}]{PerinaJr2006}
\bibinfo{author}{\bibfnamefont{J.}~\bibnamefont{{Pe\v{r}ina~Jr.}}},
  \bibinfo{author}{\bibfnamefont{M.}~\bibnamefont{Centini}},
  \bibinfo{author}{\bibfnamefont{C.}~\bibnamefont{Sibilia}},
  \bibinfo{author}{\bibfnamefont{M.}~\bibnamefont{Bertolotti}},
  \bibnamefont{and} \bibinfo{author}{\bibfnamefont{M.}~\bibnamefont{Scalora}},
  \bibinfo{journal}{Phys. Rev. A} \textbf{\bibinfo{volume}{73}},
  \bibinfo{eid}{033823} (\bibinfo{year}{2006}).

\bibitem[{\citenamefont{Li et~al.}(2005)\citenamefont{Li, Voss, Sharping, and
  Kumar}}]{Li2005}
\bibinfo{author}{\bibfnamefont{X.}~\bibnamefont{Li}},
  \bibinfo{author}{\bibfnamefont{P.~L.} \bibnamefont{Voss}},
  \bibinfo{author}{\bibfnamefont{J.~E.} \bibnamefont{Sharping}},
  \bibnamefont{and} \bibinfo{author}{\bibfnamefont{P.}~\bibnamefont{Kumar}},
  \bibinfo{journal}{Phys. Rev. Lett.} \textbf{\bibinfo{volume}{94}},
  \bibinfo{eid}{053601} (\bibinfo{year}{2005}).

\bibitem[{\citenamefont{Fulconis et~al.}(2005)\citenamefont{Fulconis, Alibart,
  Wadsworth, Russell, and Rarity}}]{Fulconis2005}
\bibinfo{author}{\bibfnamefont{J.}~\bibnamefont{Fulconis}},
  \bibinfo{author}{\bibfnamefont{O.}~\bibnamefont{Alibart}},
  \bibinfo{author}{\bibfnamefont{W.}~\bibnamefont{Wadsworth}},
  \bibinfo{author}{\bibfnamefont{P.}~\bibnamefont{Russell}}, \bibnamefont{and}
  \bibinfo{author}{\bibfnamefont{J.}~\bibnamefont{Rarity}},
  \bibinfo{journal}{Opt. Express} \textbf{\bibinfo{volume}{13}},
  \bibinfo{pages}{7572} (\bibinfo{year}{2005}).

\bibitem[{\citenamefont{Fan et~al.}(2005)\citenamefont{Fan, Migdall, and
  Wang}}]{Fan2005}
\bibinfo{author}{\bibfnamefont{J.}~\bibnamefont{Fan}},
  \bibinfo{author}{\bibfnamefont{A.}~\bibnamefont{Migdall}}, \bibnamefont{and}
  \bibinfo{author}{\bibfnamefont{L.~J.} \bibnamefont{Wang}},
  \bibinfo{journal}{Opt. Lett.} \textbf{\bibinfo{volume}{30}},
  \bibinfo{pages}{3368} (\bibinfo{year}{2005}).

\bibitem[{\citenamefont{Abolghasem et~al.}(2009)\citenamefont{Abolghasem,
  Hendrych, Shi, Torres, and Helmy}}]{Abolghasem2009}
\bibinfo{author}{\bibfnamefont{P.}~\bibnamefont{Abolghasem}},
  \bibinfo{author}{\bibfnamefont{M.}~\bibnamefont{Hendrych}},
  \bibinfo{author}{\bibfnamefont{X.}~\bibnamefont{Shi}},
  \bibinfo{author}{\bibfnamefont{J.~P.} \bibnamefont{Torres}},
  \bibnamefont{and} \bibinfo{author}{\bibfnamefont{A.~S.} \bibnamefont{Helmy}},
  \bibinfo{journal}{Opt. Lett.} \textbf{\bibinfo{volume}{34}},
  \bibinfo{pages}{2000} (\bibinfo{year}{2009}).

\bibitem[{\citenamefont{Svozil{\'i}k et~al.}(2011)\citenamefont{Svozil{\'i}k,
  Hendrych, Helmy, and Torres}}]{Svozilik2011}
\bibinfo{author}{\bibfnamefont{J.}~\bibnamefont{Svozil{\'i}k}},
  \bibinfo{author}{\bibfnamefont{M.}~\bibnamefont{Hendrych}},
  \bibinfo{author}{\bibfnamefont{A.~S.} \bibnamefont{Helmy}}, \bibnamefont{and}
  \bibinfo{author}{\bibfnamefont{J.~P.} \bibnamefont{Torres}},
  \bibinfo{journal}{Opt. Express} \textbf{\bibinfo{volume}{19}},
  \bibinfo{pages}{3115} (\bibinfo{year}{2011}).

\bibitem[{\citenamefont{Yeh}(1988)}]{Yeh1988}
\bibinfo{author}{\bibfnamefont{P.}~\bibnamefont{Yeh}},
  \emph{\bibinfo{title}{Optical Waves in Layered Media}}
  (\bibinfo{publisher}{Wiley, New York}, \bibinfo{year}{1988}).

\bibitem[{\citenamefont{{Pe\v{r}ina~Jr.}
  et~al.}(2009{\natexlab{a}})\citenamefont{{Pe\v{r}ina~Jr.}, Centini, Sibilia,
  and Bertolotti}}]{PerinaJr2009b}
\bibinfo{author}{\bibfnamefont{J.}~\bibnamefont{{Pe\v{r}ina~Jr.}}},
  \bibinfo{author}{\bibfnamefont{M.}~\bibnamefont{Centini}},
  \bibinfo{author}{\bibfnamefont{C.}~\bibnamefont{Sibilia}}, \bibnamefont{and}
  \bibinfo{author}{\bibfnamefont{M.}~\bibnamefont{Bertolotti}},
  \bibinfo{journal}{J. Russ. Laser Res.} \textbf{\bibinfo{volume}{30}},
  \bibinfo{pages}{508} (\bibinfo{year}{2009}{\natexlab{a}}).

\bibitem[{\citenamefont{{Pe\v{r}ina~Jr.}
  et~al.}(2009{\natexlab{b}})\citenamefont{{Pe\v{r}ina~Jr.}, Centini, Sibilia,
  and Bertolotti}}]{PerinaJr2009c}
\bibinfo{author}{\bibfnamefont{J.}~\bibnamefont{{Pe\v{r}ina~Jr.}}},
  \bibinfo{author}{\bibfnamefont{M.}~\bibnamefont{Centini}},
  \bibinfo{author}{\bibfnamefont{C.}~\bibnamefont{Sibilia}}, \bibnamefont{and}
  \bibinfo{author}{\bibfnamefont{M.}~\bibnamefont{Bertolotti}},
  \bibinfo{journal}{Phys. Rev. A} \textbf{\bibinfo{volume}{80}},
  \bibinfo{pages}{033844} (\bibinfo{year}{2009}{\natexlab{b}}).

\bibitem[{\citenamefont{{Pe\v{r}ina~Jr.}
  et~al.}(2007{\natexlab{a}})\citenamefont{{Pe\v{r}ina~Jr.}, Centini, Sibilia,
  Bertolotti, and Scalora}}]{PerinaJr2007b}
\bibinfo{author}{\bibfnamefont{J.}~\bibnamefont{{Pe\v{r}ina~Jr.}}},
  \bibinfo{author}{\bibfnamefont{M.}~\bibnamefont{Centini}},
  \bibinfo{author}{\bibfnamefont{C.}~\bibnamefont{Sibilia}},
  \bibinfo{author}{\bibfnamefont{M.}~\bibnamefont{Bertolotti}},
  \bibnamefont{and} \bibinfo{author}{\bibfnamefont{M.}~\bibnamefont{Scalora}},
  \bibinfo{journal}{Phys. Rev. A} \textbf{\bibinfo{volume}{75}},
  \bibinfo{eid}{013805} (\bibinfo{year}{2007}{\natexlab{a}}).

\bibitem[{\citenamefont{Vogel et~al.}(2001)\citenamefont{Vogel, Welsch, and
  Walentowicz}}]{Vogel2001}
\bibinfo{author}{\bibfnamefont{W.}~\bibnamefont{Vogel}},
  \bibinfo{author}{\bibfnamefont{D.~G.} \bibnamefont{Welsch}},
  \bibnamefont{and}
  \bibinfo{author}{\bibfnamefont{S.}~\bibnamefont{Walentowicz}},
  \emph{\bibinfo{title}{Quantum Optics}} (\bibinfo{publisher}{Wiley-VCH,
  Weinheim}, \bibinfo{year}{2001}).

\bibitem[{\citenamefont{{Pe\v{r}ina~Jr.}
  et~al.}(2009{\natexlab{c}})\citenamefont{{Pe\v{r}ina~Jr.}, Luk\v{s}, Haderka,
  and Scalora}}]{PerinaJr2009d}
\bibinfo{author}{\bibfnamefont{J.}~\bibnamefont{{Pe\v{r}ina~Jr.}}},
  \bibinfo{author}{\bibfnamefont{A.}~\bibnamefont{Luk\v{s}}},
  \bibinfo{author}{\bibfnamefont{O.}~\bibnamefont{Haderka}}, \bibnamefont{and}
  \bibinfo{author}{\bibfnamefont{M.}~\bibnamefont{Scalora}},
  \bibinfo{journal}{Phys. Rev. Lett.} \textbf{\bibinfo{volume}{103}},
  \bibinfo{pages}{063902} (\bibinfo{year}{2009}{\natexlab{c}}).

\bibitem[{\citenamefont{{Pe\v{r}ina~Jr.}
  et~al.}(2009{\natexlab{d}})\citenamefont{{Pe\v{r}ina~Jr.}, Luk\v{s}, and
  Haderka}}]{PerinaJr2009e}
\bibinfo{author}{\bibfnamefont{J.}~\bibnamefont{{Pe\v{r}ina~Jr.}}},
  \bibinfo{author}{\bibfnamefont{A.}~\bibnamefont{Luk\v{s}}}, \bibnamefont{and}
  \bibinfo{author}{\bibfnamefont{O.}~\bibnamefont{Haderka}},
  \bibinfo{journal}{Phys. Rev. A} \textbf{\bibinfo{volume}{80}},
  \bibinfo{pages}{043837} (\bibinfo{year}{2009}{\natexlab{d}}).

\bibitem[{\citenamefont{Scalora et~al.}(1997)\citenamefont{Scalora, Bloemer,
  Manka, Dowling, Bowden, Viswanathan, and Haus}}]{Scalora1997}
\bibinfo{author}{\bibfnamefont{M.}~\bibnamefont{Scalora}},
  \bibinfo{author}{\bibfnamefont{M.~J.} \bibnamefont{Bloemer}},
  \bibinfo{author}{\bibfnamefont{A.~S.} \bibnamefont{Manka}},
  \bibinfo{author}{\bibfnamefont{J.~P.} \bibnamefont{Dowling}},
  \bibinfo{author}{\bibfnamefont{C.~M.} \bibnamefont{Bowden}},
  \bibinfo{author}{\bibfnamefont{R.}~\bibnamefont{Viswanathan}},
  \bibnamefont{and} \bibinfo{author}{\bibfnamefont{J.~W.} \bibnamefont{Haus}},
  \bibinfo{journal}{Phys. Rev. A} \textbf{\bibinfo{volume}{56}},
  \bibinfo{pages}{3166} (\bibinfo{year}{1997}).

\bibitem[{\citenamefont{Hamar et~al.}(2010)\citenamefont{Hamar,
  {Pe\v{r}ina~Jr.}, Haderka, and Mich\'{a}lek}}]{Hamar2010}
\bibinfo{author}{\bibfnamefont{M.}~\bibnamefont{Hamar}},
  \bibinfo{author}{\bibfnamefont{J.}~\bibnamefont{{Pe\v{r}ina~Jr.}}},
  \bibinfo{author}{\bibfnamefont{O.}~\bibnamefont{Haderka}}, \bibnamefont{and}
  \bibinfo{author}{\bibfnamefont{V.}~\bibnamefont{Mich\'{a}lek}},
  \bibinfo{journal}{Phys. Rev. A} \textbf{\bibinfo{volume}{81}},
  \bibinfo{pages}{043827} (\bibinfo{year}{2010}).

\bibitem[{\citenamefont{{Pe\v{r}ina~Jr.}
  et~al.}(2007{\natexlab{b}})\citenamefont{{Pe\v{r}ina~Jr.}, Centini, Sibilia,
  Bertolotti, and Scalora}}]{PerinaJr2007}
\bibinfo{author}{\bibfnamefont{J.}~\bibnamefont{{Pe\v{r}ina~Jr.}}},
  \bibinfo{author}{\bibfnamefont{M.}~\bibnamefont{Centini}},
  \bibinfo{author}{\bibfnamefont{C.}~\bibnamefont{Sibilia}},
  \bibinfo{author}{\bibfnamefont{M.}~\bibnamefont{Bertolotti}},
  \bibnamefont{and} \bibinfo{author}{\bibfnamefont{M.}~\bibnamefont{Scalora}},
  in \emph{\bibinfo{booktitle}{Conference on Coherence and Quantum Optics}}
  (\bibinfo{publisher}{Optical Society of America},
  \bibinfo{year}{2007}{\natexlab{b}}), p. \bibinfo{pages}{CSuA6}.

\end{thebibliography}
\bibliographystyle{apsrev}

\end{document}